\documentclass[conference]{IEEEtran}
 
\usepackage{url}
\usepackage{cite}
\usepackage[pdftex]{graphicx}
\usepackage{amsmath}
\usepackage{array}

\usepackage{amstext,epsfig,amssymb,amsbsy}
\usepackage{mathtools}
\usepackage[output-decimal-marker={.}]{siunitx}
\usepackage{cases}

\usepackage[]{fancyhdr} %
\newcommand{\changefont}{\fontsize{9}{9}\selectfont}
\usepackage[caption=false,font=footnotesize]{subfig}

\usepackage[utf8]{inputenc}
\usepackage[english]{babel}
\usepackage{xcolor}

\usepackage{mathdots}
\usepackage{tabularx}
\usepackage{booktabs}       

\usepackage{lipsum}

\usepackage{siunitx}
\DeclareSIUnit\pu{p.u.}

\usepackage[capitalise]{cleveref}
\usepackage{bm}
\usepackage{cancel}

\usepackage{algorithm}
\usepackage[noend]{algpseudocode}

\usepackage{tikz,pgf} 
\usetikzlibrary{shapes.geometric}
\usetikzlibrary{arrows.meta}

\IEEEoverridecommandlockouts
\begin{document}
\title{Closing the Loop: A Framework for Trustworthy Machine Learning in Power Systems}

\author{\IEEEauthorblockN{Jochen Stiasny, Samuel Chevalier, Rahul Nellikkath, Brynjar Sævarsson, Spyros Chatzivasileiadis}
\IEEEauthorblockA{
Division of Power and Energy Systems\\
Department of Wind and Energy Systems\\
Technical University of Denmark\\
Kgs. Lyngby, Denmark\\
\{jbest,schev,rnelli,brysa,spchatz\}@dtu.dk}\thanks{This work is supported by the multiDC project as funded by Innovation Fund
Denmark, Grant No. 6154-00020B, by the ERC Project VeriPhIED,
funded by the European Research Council, Grant Agreement No: 949899, and by ID-EDGe, funded by Innovation Fund Denmark, Grant No. 8127-00017B.}}


%


\lhead{ACCEPTED FOR PRESENTATION IN 11TH BULK POWER SYSTEMS DYNAMICS AND CONTROL SYMPOSIUM (IREP 2022), JULY 25-30, 2022, BANFF, CANADA}
\setlength{\headheight}{22.42pt}

\maketitle
\thispagestyle{fancy}
\pagestyle{fancy}

\begin{abstract}
Deep decarbonization of the energy sector will require massive penetration of stochastic renewable energy resources and an enormous amount of grid asset coordination; this represents a challenging paradigm for the power system operators who are tasked with maintaining grid stability and security in the face of such changes. With its ability to learn from complex datasets and provide predictive solutions on fast timescales, machine learning (ML) is well-posed to help overcome these challenges as power systems transform in the coming decades. In this work, we outline five key challenges (dataset generation, data pre-processing, model training, model assessment, and model embedding) associated with building trustworthy ML models which learn from physics-based simulation data. We then demonstrate how linking together individual modules, each of which overcomes a respective challenge, at sequential stages in the machine learning pipeline can help enhance the overall performance of the training process. In particular, we implement methods that connect different elements of the learning pipeline through feedback, thus ``closing the loop" between model training, performance assessments, and re-training. We demonstrate the effectiveness of this framework, its constituent modules, and its feedback connections by learning the N-1 small-signal stability margin associated with a detailed model of a proposed North Sea Wind Power Hub system.






\end{abstract}



%
\IEEEpeerreviewmaketitle

\section{Introduction}\label{sec:introduction}

In order to meet the carbon emission reduction goals set out in the Paris Climate Agreement, governments across the world are targeting massive renewable energy integration projects. Over the next decade, a variety of renewable energy projects related to offshore energy islands~\cite{Misyris:2020_NSWPH}, GW-scale battery energy storage systems~\cite{Amin:2021}, and multi-GW-scale solar parks~\cite{KhareSaxena:2020} are set to fundamentally transform global power systems. These large-scale projects will be simultaneously complemented by the deployment of millions of distributed energy resources (DERs). The integration of these renewable energy projects into the existing bulk power grid is ultimately emblematic of the ongoing transition towards zero-carbon energy systems.

\begin{figure}
    \centering
    \includegraphics[width=1\columnwidth]{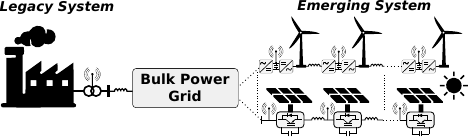}
    \caption{Legacy power grids are dominated by large, thermal synchronous generators, as shown on the left. The right side depicts the emerging paradigm, where individual thermal generators are replaced by hundreds or potentially thousands of inverter based resources (IBRs) and converter-interfaced generators (CIGs). These new devices generate massive ammounts of data and introduce intractably complex dynamics into the bulk power grid.}
    \label{fig:NSWPH_intro}
\end{figure}

As legacy grid infrastructure is enriched with these previously unseen yet massively deployed components, system operators are faced with entirely new operational challenges. Ensuring that security constraints, such as N-1 security, small-signal and voltage stability, and many others, are met in emerging grids,
especially when embedding these as constraints in optimal dispatch problems, will quickly become computationally intractable. Examples of increasing intractability, which we find throughout the emerging power grid, necessitate the development of new analysis tools. 

With its ability to efficiently learn from complex datasets, provide universal function approximation, and generate predictive solutions on  fast timescales, Machine Learning (ML) is emerging as one such tool which can help overcome intractability. Furthermore, emerging research~\cite{Kody:2021,Murzakhanov:2020} is convincingly showing that \textit{embedding} surrogate ML models directly inside safety-critical grid routines can transform intractable optimization problems into tractable ones.
The recent increase of research on ML within the power systems community is enormous, as comprehensively reviewed by~\cite{Duchesne2020}. Several works have already shown ML's potential in real-world applications~\cite{Marot:2020,Donnot:2017}; however, it is clear that ML has not yet proven ready to be widely adopted in power systems. As~\cite{Duchesne2020} elaborates on, there are many existing challenges to be addressed; one of the most critical ones is the building of \textit{trust} between domain experts and the ML modeling strategies produced by the research community. Trust will be built as the usefulness, efficiency, and reliability of ML models are tested and proven over time.

\begin{figure}[ht]
 \center
  \definecolor{colorbox}{rgb}{0.12156862745098039, 0.47058823529411764, 0.7058823529411765}
\definecolor{color_arrows}{rgb}{0.80, 0.4, 0.37}
\tikzstyle{BOXY} = [rectangle, rounded corners = 5, minimum width=3.2cm, minimum height=0.8cm,text centered, draw=black, fill=colorbox!50,line width=0.3mm, font=\small]
\def\boxdistance{0.8cm}
\begin{tikzpicture}

\node[align=center] (MLframe) [rectangle, rounded corners = 5, minimum width=8.5cm, minimum height=7.25cm,text centered, draw=black, fill=orange!20,line width=0.1mm, font=\small, yshift = -2*\boxdistance] {};


\node[align=center] (database) [BOXY, above of = MLframe, yshift = 2*\boxdistance, xshift=-1.2cm] {Dataset creation\\(Sec.~\labelcref{subsec:M_database_creation}, \labelcref{subsec:database_creation})};
\node[align=center] (inputdomain) [BOXY, above of = database, yshift = \boxdistance, xshift=-0.8cm] {Physics-based\\domain question};
\node[align=center] (dataprep) [BOXY, below of = database, yshift = -\boxdistance, xshift=0.8cm] {Data pre-processing\\(Sec.~\labelcref{subsec:M_preprocessing}, \labelcref{subsec:preprocessing})};
\node[align=center] (training) [BOXY, below of = dataprep, yshift = -\boxdistance, xshift=0.8cm] {Model learning\\(Sec.~\labelcref{subsec:M_model_training}, \labelcref{subsec:model_training})};
\node[align=center] (evaluation) [BOXY, below of = training, yshift = -\boxdistance, xshift=0.8cm] {Model assessment\\(Sec.~\labelcref{subsec:M_performance_assessment}, \labelcref{subsec:performance_assessment})};
\node[align=center] (verification) [ellipse, draw, font=\small, fill=color_arrows,  minimum width=1.5cm, minimum height=0.8cm, below of = training, yshift = -\boxdistance, xshift=-3cm] {\textbf{Reformulated}\\\textbf{model}};
\node[align=center] (embedding) [BOXY, below of = evaluation, yshift = -\boxdistance, xshift=0.8cm] {Embedding\\(Sec.~\labelcref{subsec:M_embedding}, \labelcref{subsec:embedding})};

\draw[color=black,->,>={Stealth[scale=1.2]},line width=0.3mm] (inputdomain) --  (database);
\draw[color=black,->,>={Stealth[scale=1.2]},line width=0.3mm] (database) --  (dataprep);
\draw[color=black,->,>={Stealth[scale=1.2]},line width=0.3mm] (dataprep) --  (training);
\draw[color=black,->,>={Stealth[scale=1.2]},line width=0.3mm] (training) --  (evaluation);
\draw[color=black,->,>={Stealth[scale=1.2]},line width=0.3mm] (evaluation) --  (embedding);

\draw[transform canvas={xshift=0.1cm}] (inputdomain.east) edge[bend left,color=color_arrows,->,>={Stealth[scale=1.2]},line width=0.4mm]node[midway, right, align=center] {\textbf{Physics}\\\textbf{Regularization}} (training.20);
\draw[] (dataprep.west) edge[bend left,color=color_arrows,->,>={Stealth[scale=1.2]},line width=0.4mm]node[midway, right, align=center] {\textbf{DW}} (database.south west);
\draw[] (training.west) edge[bend left,color=color_arrows,->,>={Stealth[scale=1.2]},line width=0.4mm]node[midway, left, align=center] {\textbf{NI}} (database.south west);
\draw[] (verification) edge[bend left,color=color_arrows,->,>={Stealth[scale=1.2]},line width=0.4mm]node[midway, left, align=center] {\textbf{VI}} (database.south west);
\draw[] (verification) edge[bend right,color=color_arrows,->,>={Stealth[scale=1.2]},line width=0.4mm]node[midway, left, align=center] {} (embedding.west);
\draw[color=color_arrows,->,>={Stealth[scale=1.2]},line width=0.4mm] (verification.east) -- node[midway, left, align=center] {} (evaluation.west);

\draw(training.230) edge[bend left,color=color_arrows,->,>={Stealth[scale=1.2]},line width=0.4mm] node[midway, left, align=center] {} (verification.15);

\draw[] (verification) edge[bend right,color=color_arrows,->,>={Stealth[scale=1.2]},line width=0.4mm]node[midway, left, align=center] {} (embedding.west);


\end{tikzpicture}
  \caption{Proposed ML framework, spanning from problem statement to embedding. This workflow includes feedback loops which improve training performance. These loops utilize
  Directed Walks (\textbf{DWs}), for enriching the training dataset, and Neural Network Informed (\textbf{NI}) and Verification Informed (\textbf{VI}) sampling, both of which
  use NN predictions to gather more points in high information content regions.}
  \label{fig:workflow}
\end{figure}
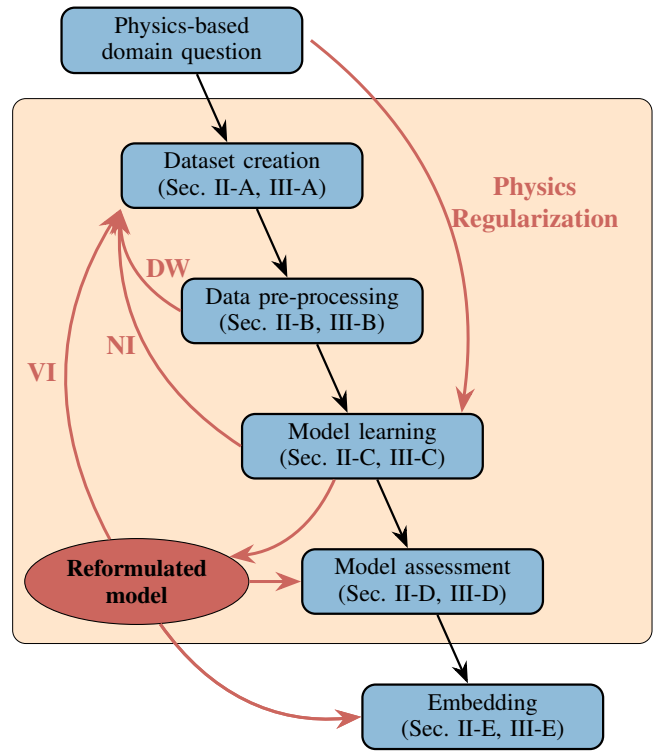

The barrier associated with building \textit{trustworthy} models naturally arises from the general complexity associated with ML. Each stage of the ML process, from data collection to model assessment, can be incredibly complex and thus pose a distinct challenge. In this paper, we consider five core challenges associated with building useful and trustworthy ML models. We focus specifically on building ML models learned from physics-based simulations in the context of safety-critical applications, e.g., security assessment and operational decision making. After stating the challenges, we then identify suitable solutions from the literature to solve these problems. The individual solutions are then organized into a coherent framework. \cref{fig:workflow} shows an overview of this framework, illustrating the various steps (subsequently called modules) that are involved in building and utilizing a ML model. For each module, substantial bodies of research already exist; some of these works are tailored for power system applications, but most are not application specific. 

Combining all of these modules into a single ML pipeline forms a process that can be difficult to handle. However, in order to produce ML models which are useful, efficient, and reliable, we believe that the following two approaches are both vital: 1) the building up of a coherent framework which considers all aspects of ML, and 2) the designing of \textit{feedback loops} within the framework in order to create synergies between the modules. \cref{fig:workflow} illustrates these two approaches by 1) considering the entire ML pipeline and 2) adding additional loops in the process signified by the red arrows to enhance model performance. The overall goal is to make each module more robust and efficient, thus enhancing the overall process itself.

The core contributions of this paper follow:
\begin{itemize}
    \item We consider five core challenges associated with using ML to learn from physics-based simulation data, and we present a coherent modular framework which utilizes state-of-the-art solutions methodologies to tackle the identified challenges.
    
    \item By utilizing feedback mechanisms, we identify how ``closing the loop" can help indicate areas of weakness within a ML model and how targeted data re-sampling and model re-training can greatly benefit the model.
    
    \item We explicitly demonstrate how the ML framework can be used to learn security-constrained stability margins from a highly detailed dynamical simulation model (i.e., a proposed North Sea Wind Power Hub system).

    \item Finally, we provide a public, open-source repository on Github \cite{Stiasny_github:2022} so that other researchers within the power systems community can test their domain-specific ML challenges using the proposed fully-integrated framework.  
    
\end{itemize}

The remainder of this paper is structured as follows. In \cref{sec:modules}, we provide an overview of five central motivating challenges in ML. In \cref{sec:framework}, we provide a literature review to identify state-of-the-art solutions for each of these challenges. We subsequently combine leading methods from the literature into a single coherent framework, and we explain how loop-closing can help link these methods together. In \cref{sec:case_study}, we present the technical details associated with the chosen case study: learning a security constrained small-signal stability margin associated with a North Sea Wind Power Hub model. We then demonstrate and compare the improved performance of the resulting models in \cref{sec:results}. In \cref{sec:discussion}, we elaborate on potential routes forward for uses of ML within the power systems research community, and we offer conclusions in \cref{sec:conclusion}.

\section{Machine learning Challenges}\label{sec:modules}
The following subsections provide an overview of the most pressing challenges associated with building trustworthy ML models. In particular, we focus on the challenges which naturally arise when deploying a ``fully integrated" ML framework, i.e., a framework that spans the full spectrum of a ML model's intended training and usage, as displayed in \cref{fig:workflow}. 
The five key topics we focus on hereafter are summarized below: 




\begin{itemize}
    \item \textit{Dataset creation}: collection of data for the modelling process, e.g., through measurements or simulation;
    \item \textit{Data pre-processing}: feature engineering of the dataset;
    \item \textit{Model training}: fitting a parameterized ML model to the provided dataset in a supervised fashion;
    \item \textit{Model assessment}: evaluation of the model performance;
    \item \textit{Embedding}: integration of the potentially reformulated model inside of its intended domain of usage.
\end{itemize}



\subsection{Dataset creation}\label{subsec:M_database_creation}


The starting point for every supervised ML algorithm is a dataset, either collected from measurements or generated from simulations. In this paper, we focus on simulation-based data collection, which presents the opportunity for creating arbitrarily large datasets. 
Practical limitations arise, however, from the computational burden of creating large and comprehensive datasets. The often cited ``curse of dimensionality" quickly renders exhaustive sampling approaches intractable\footnote{For a model with 10 input features, partitioning each feature only 10 times will lead to a search space requiring 10 billion ($10^{10}$) function evaluations.}; 
this becomes particularly pressing when the underlying physical function exhibits a high degree of non-linear behavior, or if there is a very large number of input features. As the feature number grows, the model requires a combinatorially increasing number of samples to maintain a dense coverage across the input domain. To overcome this, one may wish to avoid ``trial-and-error" sampling by directly choosing inputs which map to desirable outputs (i.e., ones close to the region of interest). Inverting nonlinear functions (e.g., time domain simulations) from output to input, however, can be highly intractable in itself.
A variety of methods have therefore been developed for more efficient sampling. With such methods, particular effort has been devoted to (i) excluding non-informative regions of the input domain on one hand, and (ii) targeting highly-informative regions more intensely on the other. Finally, for many power system applications, the initial sampling space is often unbalanced~\cite{Venzke:2021}, with a large volume of input points mapping to outputs which are infeasible or far from the region of training interest~\cite{Thams2020} (e.g., when learning a stability boundary).


\subsection{Dataset pre-processing}\label{subsec:M_preprocessing}

Before starting to learn a ML model, a pre-processing shall ensure that the subsequent learning task is well posed. This primarily relates to understanding and potentially exploiting or countering statistical properties of the dataset. The first step entails a data exploration in which statistical properties such as class imbalance or differences of scales should be identified. By subsequently applying, for example, transformations or feature selection and engineering, the dataset can be brought in a more suitable form. Another conclusion can be the need for an enrichment of the dataset. The final step of the pre-processing is to split the dataset into subsets for training, validation, and testing for the subsequent process. It is important that previously identified statistical characteristics are captured well in each of the three subsets, as the learning and assessment tasks heavily rely on their distributional similarity in order to avoid biases in the process. Only then can the ML models achieve good and reliable performance when deployed in real world application.




\subsection{Model training}\label{subsec:M_model_training}


To describe a dataset through a model, one must find some function $\bm{g}(\cdot)$ that maps the input features $\bm{x}$ to predicted outputs $\hat{\bm{y}}_{\bm{x}}$. The function itself can take a variety of different forms, but its predictions depend on a set of parameters $\theta$. The objective is to minimize the difference between the ground truth $\bm{y}_{\bm{x}}$ and prediction by adjusting the parameters $\theta$:
\begin{align}
    \min_{\theta} \left\Vert \bm{y}_{\bm{x}}-\hat{\bm{y}}_{\bm{x}}\right\Vert.
\end{align}
This formulation is equally valid for a simple linear regression model and for a highly complex deep neural network (NN); however, such models differ significantly in the degree of complexity they can represent. Finding an adequate level of representation capacity and a suitable set of parameters are the core problems associated with training a NN. An overly simplistic model will capture general trends, but it will miss more detailed aspects of the underlying mapping (``underfitting"). An overly complex model, however, might perform well on training data but poorly on unseen testing data (``overfitting"). 
To balance this trade-off and control the influence of other parameters such as optimizer settings, or regularization strengths, so-called ``hyper-parameter tuning" is conducted. It serves as a model selection process by judging the model performance on the validation dataset. This model selection can require many repeated training runs to identify well-performing setups, and it is essential for ensuring repeatability and robustness. Achieving these two key properties can enable fair comparisons between different ML models.

\subsection{Performance assessment}\label{subsec:M_performance_assessment}



Data handling practices strongly influence the performance assessment of ML models. If the statistics of the test dataset are known to match the ones of the process we intend to model, we can formulate a valid expectation on the model performance. If the underlying statistics do not match (i.e., the model is deployed ``out of sample"), then the trained model can achieve an excellent performance on the test dataset but perform very poorly when deployed. This leads to a mistaken overtrusting of the model.
In either case, this assessment is a purely statistical quantity, and it offers no indication of the model's expected behavior at unseen points. This can lead to surprisingly bad predictions and unexpected behavior. Accordingly, such statistical methods cannot offer rigorous guarantees for NN performance across the full input domain. In safety-critical applications, stronger guarantees, such as Neural Network verification, are often desired, potentially across the global operating region. This is one of the topics we will discuss in this paper.

\subsection{Model embedding}\label{subsec:M_embedding}

Typically, the goal of deploying a trained ML model serves either (i) solving a previously unsolvable or intractable problem, or (ii) estimating a solution much faster than established methods can. The latter can be especially helpful for screening scenarios where evaluation speed is critical. One challenge, however, is that ML models do not necessarily offer insights as to why a certain solution was predicted. Another general challenge is associated with ``embedding'' the ML model inside of a broader optimization or control problem. Embedding allows ML models to be flexibly utilized in a diversity of contexts. Such embedding, however, required finding suitable representations to incorporate the NN's predictions into such problems, since NNs in their common form are not well-suited to be solved via standard numerical solvers (e.g., Newton-based methods). A final challenge relates to ensuring the continued applicability of the learned model in its intended real world usage and taking actions in case problem characteristics changed since the ML model was trained and assessed.




\section{Overcoming Machine Learning Challenges: Integrating Solutions into a Workflow}\label{sec:framework}
In the following subsections, we review state-of-the-art solutions from the literature associated with the main challenges of the five modules outlined in \cref{sec:modules}. We subsequently organize selected solutions into a modular workflow, and we identify the feedback loops which can create productive synergies between the modules. First, though, we introduce useful mathematical notation to properly describe the modeling challenges.

\textit{Mapping problem:} We define a physical power system model $\mathcal{M}$ whose mapping of interest $\bm{f}:\mathbb{R}^n \xrightarrow{} \mathbb{R}^m$ is
\begin{align}\label{eq: M}
    \mathcal{M}: \bm{y}_{\bm{x}} &= \bm{f}(\bm{x}),
    \end{align}
with input domain $\mathcal{X}: \bm{x} \in \mathbb{R}^n$ and output range $\mathcal{Y}: \bm{y}_{\bm{x}} \in \mathbb{R}^m$. Usually, we are interested in learning the mapping only on some bounded input domain: $\underline{\bm{x}} \leq \bm{x} \leq \overline{\bm{x}}$. We assume there is some subset of the output space that has a disproportionately ``high information content"~\cite{Genc:2010}; this could be near a stability boundary. We denote this output region as ${\mathcal Y}^*\subseteq{\mathcal Y}$.


\textit{Neural network:}
We generally refer to the NN mapping by
\begin{align}
    \hat{\bm{y}}_{\bm{x}} = {\rm NN}(\bm{x}).
\end{align}
The function ${\rm NN}(\cdot)$ could take many forms, but we assume in the following a simple feed-forward network, with ReLU (Rectifier Linear Unit) activation functions, formulated as 
\begin{alignat}{2}
     \bm{z}_0 &= \bm{x} &&\label{eq:NN_input}\\
    \hat{\bm{z}}_{k+1} &= \bm{W}_{k+1} \bm{z}_k + \bm{b}_{k+1}, && \forall k = 0, 1, \ldots, K-1\label{eq:NN_hidden_layers}\\
    \bm{z}_k &= \max (\hat{\bm{z}}_k, 0), && \forall k = 1, \ldots, K\label{eq:ReLUactivation} \\
    \hat{\bm{y}}_{\bm{x}} &= \bm{W}_{K} \bm{z}_K + \bm{b}_{K}, \label{eq:NN_output}
\end{alignat}
where $K$ represents the number of layers, each of which is defined by a weight matrix $\bm{W}_k$ and a bias vector $\bm{b}_k$. Their sizes are $[N_{k+1} \times N_k]$ and $[N_{k+1} \times 1]$ respectively, where $N_k$ is the number of neurons in the k-th layer. The NN training problem then reads as
\begin{align}
    \min_{\bm{W}_k,\bm{b}_k} & \left\Vert \bm{y}_{\bm{x}}-\hat{\bm{y}}_{\bm{x}}\right\Vert\\
    \text{s.t.} \quad & \eqref{eq:NN_input} - \eqref{eq:NN_output}.
\end{align}

\textit{Mixed Integer Linear Program (MILP) reformulation:} When using ReLU activation functions (as shown by \eqref{eq:ReLUactivation} in our case), it is possible to  reformulate \eqref{eq:NN_input}-\eqref{eq:NN_output} as a MILP. It hinges around replacing the ReLU activation function by a set of linear constraints and $N_k$ binary variables \cite{Vincent:2017}:
\begin{subnumcases}
{\bm{z}_k = \max(\hat{\bm{z}}_k,0)\Rightarrow}
\bm{z}_k  \leq \hat{\bm{z}}_k - \hat{\bm{z}}^{\text{min}}_k  (1-\mathbf{b}_k) \label{Eq:Milp1} \\ 
\bm{z}_k  \geq \hat{\bm{z}}_k \label{Eq:Milp2}   \\
\bm{z}_k  \leq \hat{\bm{z}}^{\text{max}}_k \mathbf{b}_k  \label{Eq:Milp3}  \\
\bm{z}_k   \geq \bm{0}  \label{Eq:Milp4}  \\
\mathbf{b}_k \in \{0,1\}^{N_k} \label{Eq:Milp5}.
\end{subnumcases}
This exact reformulation of the NN prediction to a MILP allows us to integrate it to any optimization or control problem, which we can exploit, e.g., for the model assessment and embedding modules.

\subsection{Dataset creation and enrichment}\label{subsec:database_creation}



A number of classical ML textbooks (e.g.,~\cite{Goodfellow:2016}) provide a review of data collection practices which are rigorously rooted in statistical sampling theorems. In this module, we assume a dataset can be created by querying a physics-based model $\mathcal{M}$. That is, we can arbitrarily sample in $\mathcal{X}\subseteq\mathbb{R}^n$ to produce a data point $\bm{y}_{\bm{x}}$ from (\ref{eq: M}). In \ref{ss: sampling_lit}, we first present an overview of methods from the literature used for producing representative datasets from physics-based power system models. In \ref{eq: ss_DW}-\ref{eq: ss_VI}, we then present ideas concerning how to ``enrich" the dataset, either based on directed walks (DWs), or by using the NN itself to recommend targeted regions of retraining.

\subsubsection{Sampling approaches in the power systems literature}\label{ss: sampling_lit}
Using tools from mathematical finance,~\cite{Hamon:2016} employs importance sampling to generate probabilistic security assessment risk data. Importance sampling is also employed in~\cite{Krishnan:2011} for decision tree training, but a two stage procedure is employed in order to bias the sampling routine towards the high information content input space. 

To sample quickly from the feasible power flow space,~\cite{Molzahn:2017} proposed a hypersphere ``grid pruning'' algorithm rooted in SOCP and Lasserre hierarchies. This work was extended in~\cite{Thams2020}, where SDP-based hyperspheres are constructed in order to classify regions of infeasibility, thus enabling rapid rejection sampling of future input points; this work found power flow solutions that were both feasible and N-1 secure. To improve upon these methods, authors in~\cite{Venzke:2021} utilize the quadratic convex (QC) relaxation~\cite{Coffrin:2016} of the power flow problem in order to generate infeasibility certificates based on separating hyperplanes (rather than hypershperes). Hit and run sampling~\cite{Kaufman:1998}, a Monte Carlo sampling approach for generating asymptotically distributed datasets, has also found application in the power systems literature. Most recently, the developers of the OPF-Learn tool~\cite{Jones:2021} use hit and run sampling to ``uniformly and quickly'' sample high dimensional polytopes. Using SOCP relaxations, separating hyperplanes are then constructed for infeasibility classification. 


Other works related to dynamic and static security assessment have focused more sharply on collecting targeted datasets close to a pre-defined stability boundary (e.g., a small-signal~\cite{Thams2020}, transient~\cite{Wehenkel:1994}, or steady state~\cite{Hatziargyriou:1994} boundary) or addressing dataset imbalances by including the associated risk and cost of a data point in the model learning and assessment \cite{cremer_machine-learning_2021}. In~\cite{Genc:2010}, a dynamic security assessment dataset is iteratively enriched through the addition of points which are at the midpoint between secure and insecure operating points (OPs). To improve upon this method,~\cite{Thams2020} utilized a so-called ``directed walk'' method which iteratively pushed operating points close to a small-signal stability boundary via gradient descent.

\subsubsection{Enhancing datasets with Directed Walks (DWs)}\label{eq: ss_DW}

When pre-processing datasets for learning, it can become clear that the statistical properties of the dataset are not satisfactory, e.g., due to an insufficient representation of critical areas in the output domain. Therefore, one of the approaches we utilize in this paper is the directed walk methodology developed in~\cite{Thams2020} in order to target data points which lie close to a pre-defined stability threshold. 

\subsubsection{Enhancing datasets with neural network informed (NI) points}\label{eq: ss_NI}
The NN itself, even in a not-fully trained state, can provide an indication of which points in the input domain map to regions in the output domain that are most critical for obtaining a balanced and informative dataset. These mappings might not be perfectly accurate, but they can be used to perform a pre-selection of points that are first re-evaluated using the physical model and then added to the dataset. \Cref{algo:NI}, which is first proposed in this paper, describes the essential steps of this approach, which can be viewed as a practical way to close the loop between model training, NN prediction, and model retraining.

\begin{algorithm}
\caption{Network Informed (NI) dataset enrichment}

{\small \textbf{Require:}
NN, target region $\mathcal{R}_c$, $N_{\text{samples}}$ 

\begin{algorithmic}[1]\label{algo:NI}

\State Create a large number (e.g., 1'000'000) of random samples $\bm{x}$ in the input domain $\mathcal{X}$

\State Evaluate neural network $\hat{\bm{y}}_{\bm{x}} = {\rm NN}(\bm{x})$

\State Select all samples $\bm{x}$ with $\hat{\bm{y}}_{\bm{x}}\in\mathcal{R}_c \xrightarrow{} \bm{x}_c$

\State Select $N_{\text{samples}}$ NI samples out of $\bm{x}_c \xrightarrow{} \bm{x}_{NI}$

\State Evaluate ground truth $\bm{y}_{\bm{x}, NI} = \bm{f}(\bm{x}_{NI})$

\State \Return NI dataset ($\bm{x}_{NI}$, $\bm{y}_{\bm{x}, NI}$)

\end{algorithmic}}
\end{algorithm}

\subsubsection{Enhancing datasets using verification informed (VI) points}\label{eq: ss_VI}
This approach is also proposed for the first time in this paper, and it also aims to ``close the loop" (see~\cref{fig:workflow}). Compared to the NI approach, instead of simply choosing sampled data based on NN
evaluations, we now evaluate entire regions of the NN input domain by searching for certain predictions. By utilizing the NN's exact MILP reformulation from \eqref{Eq:Milp1}-\eqref{Eq:Milp5}, it is possible to pose an optimization problem that allows us to test entire regions of the NN for certain behavior (see, e.g.,~\cite{Venzke:2020}, where the MILP reformulation is used to extract worst-case performance guarantees). With this reformulation, we can use optimization-based tools to efficiently determine regions in the input domain that either are, or are not, of interest, and we can find so-called adversarial points (i.e., points whose NN predictions deviate sharply from other points close by). By feeding this information back into the dataset creation module, we can enrich the dataset with more targeted re-sampled data points. \cref{algo:VI} illustrates the procedure, where we search for values of $\epsilon$ which certify particular regions of NN prediction (see, e.g., \eqref{eq: MILP_verification} and the associated illustration in \cref{fig:Verification}).


\begin{algorithm}
\caption{Verification Informed (VI) dataset enrichment}

{\small \textbf{Require:}
NN, target region $\mathcal{R}_c$, $N_{\text{samples}}$ 

\begin{algorithmic}[1]\label{algo:VI}

\State Create a large number (e.g., 1'000'000) of random samples $\bm{x}$ in the input domain $\mathcal{X}$

\For{each reference point $\bm{x}_0$ }

\State Find $\epsilon_i$ which verifies neural network up to the border of $\mathcal{R}_c$  

\State Define region non-interest: $||\bm{x} - \bm{x}_0||\le\epsilon_i$  $\xrightarrow{} \mathcal{R}_i$


\EndFor {\bf end}

\State Select all samples $\{\bm{x}\notin{\mathcal R}_i,{\forall i}\}\xrightarrow{} \bm{x}_c$


\State Select $N_{\text{samples}}$ VI samples out of $\bm{x}_c \xrightarrow{} \bm{x}_{VI}$

\State Evaluate ground truth $\bm{y}_{\bm{x}, VI} = \bm{f}(\bm{x}_{VI})$

\State \Return VI dataset ($\bm{x}_{VI}$, $\bm{y}_{\bm{x}, VI}$)

\end{algorithmic}}
\end{algorithm}

\subsection{Dataset pre-processing}\label{subsec:preprocessing}

The focus of pre-processing the dataset lies in preparing the dataset in such a way that the subsequent training problem can be solved efficiently. This step can range from applying straightforward best practices to highly involved techniques; in any case, having a profound understanding of the present dataset can prove vital for all subsequent steps. Therefore, it is worth combining pre-processing with data exploration techniques so one can be aware of potential statistical particularities of a dataset.

In terms of practical aspects of pre-processing, the training can usually be aided by standardizing the input data; ideally, the input features are also uncorrelated \cite{montavon_neural_2012}. The reasons behind this relate to the optimization algorithms used in the training process performing more efficiently. Furthermore, standardization helps to avoid unintended effects due to scaling when using regularization in the training. A more fundamental and theoretical question relates to the selection or engineering of input and output features, i.e., the question of which features or combinations of features in the dataset are the most important ones for accurate predictions. The idea of ``engineering'' the learning task links to the assumption that the mapping which we attempt to learn could be expressed in a lower dimension if an appropriate set of coordinates, or more formally, the appropriate manifold, is identified. Selecting and engineering input and output features plays an important role for subsequent steps of the process, and it can greatly improve of hurt the chances of success. 

Refs.~\cite{hastie_elements_2009, blum_selection_1997, guyon_introduction_2003} describe relevant techniques in a general ML context, for reviews for power systems we refer to \cite{Duchesne2020, salcedo-sanz_feature_2018}.
In a setup where a dataset is generated from simulations, using the available domain knowledge can render an advanced feature selection process mostly unnecessary; nonetheless, a different representation of the data is worth considering to ease the learning task.
Pre-processing can also include the ``cleaning'' of a dataset. This could entail, among other things, removing or imputing missing values or properly treating of outliers. This delicate procedure is beyond the scope of this work, since we assume a simulation-based dataset creation. 
However, for measurement-based datasets this ``cleaning'' is critical.

At the stage of the pre-processing it is also advisable to decide on the methodology of model training, model selection and model assessment. If the base dataset is sufficiently large in terms of data points, the preferred set-up is to perform a random three-fold split of the dataset into training, validation, and testing subsets. Common practice is to retain 20\% of the dataset as the hold-out or test dataset to obtain an unbiased estimation on the generalization error. Quoting Hastie et al. \cite{hastie_elements_2009} for a best practice: \textit{``Ideally, the test set should be kept in a `vault', and be brought out only at the end of the data analysis.''} Data analysis, in this context, corresponds to the model assessment, i.e., the model that seems to describe the analyzed data ``best''. The remaining dataset is then further split, for example, at a ratio 80\% to 20\% for the training and validation set. These numbers should not be taken too literally, but rather act as guides, since they can be problem and setup specific. The respective sections in \cite{Goodfellow:2016} and in particular \cite{hastie_elements_2009} provide a good introduction into the underlying concepts and statistical considerations for selecting splitting ratios. The critical aspect in splitting is that the three subsets are distributionally as similar as possible. Otherwise, the model selection and generalization error estimation can be severely biased. For setups with fewer data points, \cite{hastie_elements_2009} also elaborates on resampling methods such as cross-validation and bootstrapping techniques. Instead of random splitting, advanced tools, such as Algorithmic Splitting~\cite{Kahloot:2021}, are also being developed to determine superior dataset splitting for particular training problems. For setups with access to the data generating process, i.e., the underlying physical model for creating data points, usually the three-fold split is possible.



\subsection{Model training and physics-based regularization}\label{subsec:model_training}

In its vanilla form, the training process is guided purely by the objective of minimizing the loss function $\mathcal{L}$ given by
\begin{align}
    \mathcal{L}=\left\Vert \bm{y}_{\bm{x}}-\hat{\bm{y}}_{\bm{x}}\right\Vert,
\end{align}
where $\bm{y}_{\bm{x}}$ stems from a ground truth dataset, and $\hat{\bm{y}}_{\bm{x}}$ represents the NN's prediction. Even though it often possible to drive the loss value on the training dataset close to zero, the generalization error, i.e., a measure of the performance on the testing dataset, can turn out to be very high. This describes the phenomenon of ``over-fitting'' which we try to avoid by controlling the generalization error on the validation set during the training. This can take the form of ``early stopping'' as explained in \cite{montavon_neural_2012}, where the training is stopped as soon as the the error on the validation set does not improve significantly anymore. Another method on the level of the training is to introduce regularization, such as applying 1-norm or 2-norm regularization to the weights of the NN (i.e., penalizing the magnitudes of the weights). Regularization influences the bias-variance-trade-off of the models \cite{hastie_elements_2009}, and if its strength is well-tuned, it helps the training algorithm generate models which are both more generalizable and more parsimonious~\cite{Brunton:2019}. This leads directly to a third area of managing generalization error: ``hyper-parameter tuning''. This refers to the problem of finding suitable parameters for the training process, e.g., optimizer settings or regularization strength, and for the parameters governing the model complexity, e.g., the size of the NN. Since these hyper-parameters influence the model performance in terms of generalization capacity in non-trivial ways, one usually varies them either by random trials, a structured grid search, or by relying on a Bayesian approach. The validation dataset serves once more as a check for the generalization capacity and is used for model selection, where the term ``model" also incorporates the training hyper-parameters.

As previously mentioned, regularization is a powerful tool for increasing generalization capacity. Since power system problems often fall into domains governed by physical models, the addition of physics-based regularization might be possible and can potentially offer performance improvements. To add this regularization, the loss function $\mathcal{L}$ is, for example, extended by a comparison between the derivatives of the physical mapping and NN mapping, as in \eqref{eq:loss_regularized}. More generally, regularization can be provided through a system-specific function $\bm{h}(\cdot)$ which links the inputs, outputs, and potentially their derivatives. Thereby, differential and algebraic relations among the variables of the governing model can be incorporated and deviating predictions get penalized:
\begin{equation}\label{eq:loss_regularized}
\begin{split}
\mathcal{L}' & =\left\Vert \bm{y}_{\bm{x}}-\hat{\bm{y}}_{\bm{x}}\right\Vert+\alpha\left\Vert \frac{\partial\bm{y}_{\bm{x}}}{\partial\bm{x}}-\frac{\partial\hat{\bm{y}}_{\bm{x}}}{\partial\bm{x}}\right\Vert \\&+\beta\left\Vert \bm{h}\left(\bm{x},\hat{\bm{y}}_{\bm{x}},\frac{\partial\hat{\bm{y}}_{\bm{x}}}{\partial\bm{x}}\right)\right\Vert.\end{split}
\end{equation}
These additional terms can be weighted by $\alpha$ and $\beta$, and they can be used flexibly depending on physical model availability. The NN's derivative can usually be computed using so-called automatic differentiation \cite{baydin_automatic_2018}; for the physical model, gradients can either be derived directly from the model formulation, e.g., for differential equations, or they can be evaluated numerically.

\subsubsection{Regularization approaches in power systems}
Many recent results have focused on a particular regularization architecture, known as Physics Informed Neural Networks (PINNs)~\cite{Raissi:2019}, where partial/ordinary differential equation (P/ODE) functions explicitly regularize NN training. While this technique is relatively new, it has quickly attracted thousands of followup extensions and associated publications~\cite{Cuomo:2022}.

PINNs were first applied to power systems in~\cite{Misyris:2020} to predict swing equation dynamics. They have since additionally been used for system identification~\cite{Stiasny:2021}, transient stability predictions~\cite{stiasny_transient_2021}, and for learning grid dynamics without simulation data~\cite{Stiasny:2021_Learning}. Beyond ODE simulation and trajectory prediction, physics and sensitivity informed methods have also been utilized for regularizing models related to power distribution grid optimization~\cite{Singh:2021}, ACOPF~\cite{Singh:2020,Nellikkath2:2021}, DCOPF~\cite{Nellikkath:2021}, parameter estimation~\cite{Pagnier:2021}, and risk-aware voltage
optimization via ``risk-regularization"~\cite{Lin:2021}.

In this paper, we utilize numerically computed gradient information in order to directly regularize the training process. Incidentally, much of this gradient information is ``left over" from the directed walk dataset enrichment step, so it efficiently serves as a double use.


\subsection{Neural network verification and performance assessment}\label{subsec:performance_assessment}

Verifying the performance of NN-based ML models has become an increasingly important subfield of AI research; an excellent overview is provided in~\cite{Xiang:2018}. In this subfield, researchers have predominantly focused in the past on quantifying the so-called adversarial robustness of a NN~\cite{Xiang:2018}, other approaches attempt to quantify uncertainty through algorithms using conformal predictions as introduced by Vovk et al. \cite{vovk_algorithmic_2005} and extended in \cite{tibshirani_conformal_2019}; such methods, however, cannot offer rigorous guarantees for NN performance. In safety-critical applications, stronger guarantees are often desired, potentially across the global operating region. 

To generate provable robustness and worst-case performance guarantees, researchers have developed methods which transform the NN mapping into an equivalent set of optimization-based constraints. Such methods have utilized convex relaxations of activation functions and Lagrangians, norm-bounding, and abstract interpretations~\cite{Wong:2018,Dvijotham:2018,Dathathri:2020,Fazlyab:2020,Dvijotham:2020,Mirman:2018,Gehr:2018,Weng:2018,Zhang:2018,Hein:2017,Wang3:2018}. In the field of Reinforcement Learning (RL), researchers have also focused on training agents which are safe by design (i.e., “safe RL”)~\cite{Garcia:2015}, via e.g., learning control policies based on Lyapunov design principles~\cite{Berkenkamp:2017} and reachability analysis~\cite{Gillula:2012}. Such methods can suffer from (i) learning suboptimal control policies and (ii) an inability to provide guarantees against all possible contingencies~\cite{Xiang:2018}.


An alternative verification approach relies on modelling the ReLU-based NN as a binary MILP~\cite{Dutta:2018,Lomuscio:2017,Vincent:2017,Xiao:2018}. Unrelaxed MILP verification is known as a “complete verification algorithm” in the literature, because it utilizes an exhaustive search of the constrained input space and therefore can produce exact, worst-case guarantees~\cite{Dathathri:2020}. While MILPs are generally NP-hard to solve, MILP solvers have become tremendously powerful over the last decade. With light sparsity regularization and pruning, the authors in~\cite{Xiao:2018} could solve verification problems on NNs with 1000+ ReLUs in 10s of seconds.



Power system operation and control problems are considered safety-critical applications; therefore, ML models should meet strong verification standards if they are ever to be realistically deployed within these realms. Complete verification methods (reviewed above) can generate the strongest and tightest possible guarantees across large operational possibilities and are, therefore, considered a promising research direction. This approach was first applied to power systems in~\cite{Venzke2:2020}, where the authors computed formal guarantees for the performance of classification NNs in the context of security assessment. This method was extended in~\cite{Venzke:2020} and~\cite{Nellikkath2:2021}, where the authors computed worst-case performance guarantees for a regression NN which was trained to solve the DC- and AC-OPF problems, respectively. A feedback procedure is developed in~\cite{Nellikkath:2021}, which finds the worst-case performing input point, and then adds this points (and its true ground-truth output) into the training set.

In this paper, we assume access to the underlying physical model (i.e., the dynamic power system model describing the North Sea Wind Power Hub) for training purposes; however, the mapping from input (power dispatch and control parameters) to output (worst-case security constrained damping ratio) is overly intractable to be used for generating global worst-case performance guarantees. Accordingly, we explore alternative MILP-based verification approaches which iteratively (i) locate regions where training data is lacking and then (ii) solve a bounded worse-case performance problem in order to add targeted points back into the training dataset.


\subsection{Embedding neural networks in optimization problems}\label{subsec:embedding}
Increasingly, ML models have been deployed as surrogate models which replace intractable constraints inside of challenging optimization problems. Generally, this procedure is referred to as ``embedding" a ML model inside of an optimization problem~\cite{Grimstad:2019}, and it has enjoyed a growing spectrum of uses, e.g., via MILP reformulation of multivariate adaptive regression splines~\cite{Martinez:2017}, co-prediction and optimization of unknown parameter values~\cite{Elmachtoub:2021}, and optimization over tree ensembles~\cite{Misic:2017}. In particular, reformulating ReLU-based NNs into equivalent MILP constraints~\cite{Vincent:2017} is becoming a popular tool for embedding NNs inside of optimization problems~\cite{Huchette:2020,Say:2017,Katz:2020}. While this approach is highly flexible, the computational expense of solving the associated MILP grows quickly with NN size, becoming nearly intractable for NNs with thousands of nodes~\cite{Grimstad:2019}.

NN embedding has only been applied within the power systems research community in few applications so far. Authors in~\cite{Zhang:2020} approximate the nonlinear function between microgrid operating point and frequency nadir in order to encode frequency constraints in a scheduling problem. Surrogate frequency constraints are then encoded into the linearized Unit Commitment problem in~\cite{Zhang:2021}. In order to solve an AC Unit Commitment problem,~\cite{Kody:2021} replaces the AC power flow equations with a piecewise linear approximation learned by an optimally compact NN. In~\cite{Murzakhanov:2020}, a NN learns to classify regions of small-signal stability in order to solve a security constrained OPF problem. Input convex NNs, as popularized in~\cite{Amos:2016}, are used in~\cite{Chen:2020} to learn a reactive power control law for optimal voltage regulation.~\cite{Misyris:2021} learns grid-following converter dynamics with a NN, and then uses a MILP reformulation to estimate associated critical clearing times.

This paper introduces the benefits that trustworthy ML can offer in power systems, and it demonstrates them on offshore wind hubs in order to support planners and operators who must safely and optimally dispatch their systems. Accordingly, after building a ReLU-based NN which maps from dispatch decisions and control parameters to damping ratios, we embed this mapping inside of an illustrative security-constrained dispatch problem. This embedding process utilizes the MILP reformulation proposed in~\cite{Vincent:2017}, and it highlights the capacity for a NN to learn intractably complex constraint mappings.

\subsection{Brief overview of the full framework}
We now summarize the core modules in the proposed framework, as depicted by the workflow in \cref{fig:workflow}. 

\textit{Usage criteria:} This framework is \textit{most applicable} for building surrogate ML models from physics-based simulation data whose underlying input/output mapping can be arbitrarily (although, potentially expensively) queried. Furthermore, we assume that naively generated training datasets will be unbalanced (i.e., they will not adequately represent the regions of highest interest), and we therefore need to develop approaches to address that. Finally, we assume the resulting ML model will be embedded inside of an optimization or control problem.

\textit{Framework summary:} We begin by defining bounded input $\mathcal X$ and output $\mathcal Y$ domains of interest associated with a physical mapping \eqref{eq: M}. In particular, the set of output features are chosen to be sufficiently expressive such that they can be used to solve the targeted application (embedding) problem. We then sample from the input space ${\mathcal X}$ using a statistically grounded sampling method (e.g., Latin Hypercube sampling). For output points which map into the (pre-defined) high information content region ${\mathcal Y}^*$, we numerically approximate the associated gradient information (e.g., via \eqref{eq: sensitivity}), and we apply the directed walk method (e.g., via \eqref{eq: SD}) to enrich the training dataset. Gradients associated with all data are then computed in order to regularize the training process via (\ref{eq:loss_regularized}). 

Before training, the dataset (with the directed walk points) is split into training, validation, and testing subsets and the input and output datasets are standardized. 
Then the training is conducted using the validation dataset as a monitoring of the generalization error. For the ``loop-closing" processes the training is halted at an intermediate stage and one of the two subsequent resampling procedures is initiated. First, a large set of new input samples are chosen and rapidly evaluated with the NN. Points which map into $\hat{\mathcal Y}^*$ are saved and evaluated by the ground truth mapping \eqref{eq: M}; we refer to this process as \textit{Neural Network Informed} (NI) sampling. Second, we reformulate the (ReLU-based) NN as the MILP via \eqref{eq:NN_input}, \eqref{eq:NN_hidden_layers}, \eqref{eq:NN_output}, \eqref{Eq:Milp1}-\eqref{Eq:Milp5}. The resulting MILP is used to find both adversarial points (i.e., points where small perturbation lead to large prediction changes) and points whose predictions maximally deviate from some expectation (see \cref{ss:assessment_verification} for examples and details). We refer to this process as \textit{Verification Informed} (VI) sampling. All model training undergoes a hyper-parameter tuning before assessing the performance on the final test dataset. Lastly, the MILP reformulation is used to embed the now-trustworthy model inside a relevant problem (see \cref{ss:embeding_NSWPH} for examples and details).

\section{Case study: The North Sea Wind Power Hub}\label{sec:case_study}

\begin{figure}
 \center
 \includegraphics[width=1\columnwidth]{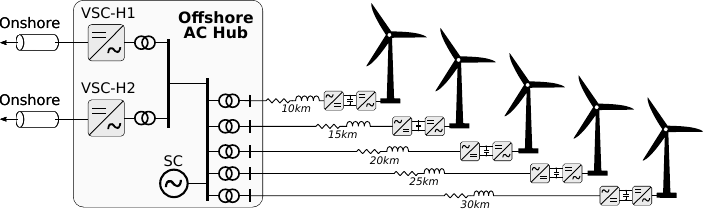}
  \caption{The North Sea Wind Power Hub. The offshore AC hub connects to five wind turbines, and a local synchronous condenser (SC) that offers inertia and voltage support. Hub voltage source converters (VSCs) supply DC power to onshore (mainland) power systems via HVDC sub-sea cables. The associated dynamical system model contains 164 differential states.}
  \label{fig:NSWPH}
\end{figure}

In this section, we present a proposed model of the North Sea Wind Power Hub (NSWPH). This model represents a realistic example of an emerging converter-dominated system that must integrate securely into the legacy bulk power grid. Converter-dominated systems have many more degrees of freedom compared to conventional power systems, and they also require several additional control loops to maintain a stable operation. Tuning all necessary control parameters to ensure stability and achieve a good performance becomes an increasingly complex task; it requires advanced optimization tools to explicitly consider the stability constraints of the system. This has so far been very difficult to achieve with conventional optimization tools. 
This paper proposes a machine learning framework to overcome this otherwise intractable optimization problem. 
The objective is to optimally tune the control parameters and determine the power dispatch of the NSWPH system under small-signal stability and N-1 security constraints. \Cref{fig:NSWPH} depicts a circuit diagram of a proposed North Sea Wind Power Hub (NSWPH), with 5 wind farms, a synchronous condenser, and two HVDC Voltage Source Converters (VSCs) that connect the offshore AC Hub to the mainland. In modeling this system, we focus exclusively on the dynamics of the offshore AC grid (Hub) that includes the synchronous connection between the wind turbine VSCs, the hub VSCs and the synchronous condenser. The technical details associated with the NSWPH system and VSCs are reviewed in~\cref{App_NSWPH} and in~\cite{Misyris:2020_NSWPH,Bastin:2019}.

When planning the dispatch of this system, operators are assumed to have control over the active and reactive power set points of the turbines ($P_{\rm ref}$ and $Q_{\rm ref}$, respectively). In future power systems, operators may also depend on the wind hubs to provide both primary frequency and primary voltage control support services. Accordingly, we assume operators have the capacity to adjust the droop parameters which determine the system's participation in both primary frequency and primary voltage control ($K_{p,f}$ and $K_v$, respectively). These parameters are depicted inside the active and reactive power control loops of the VSCs in \cref{fig:NSWPH_Control}.

\begin{figure}
 \center
 \includegraphics[width=1\columnwidth]{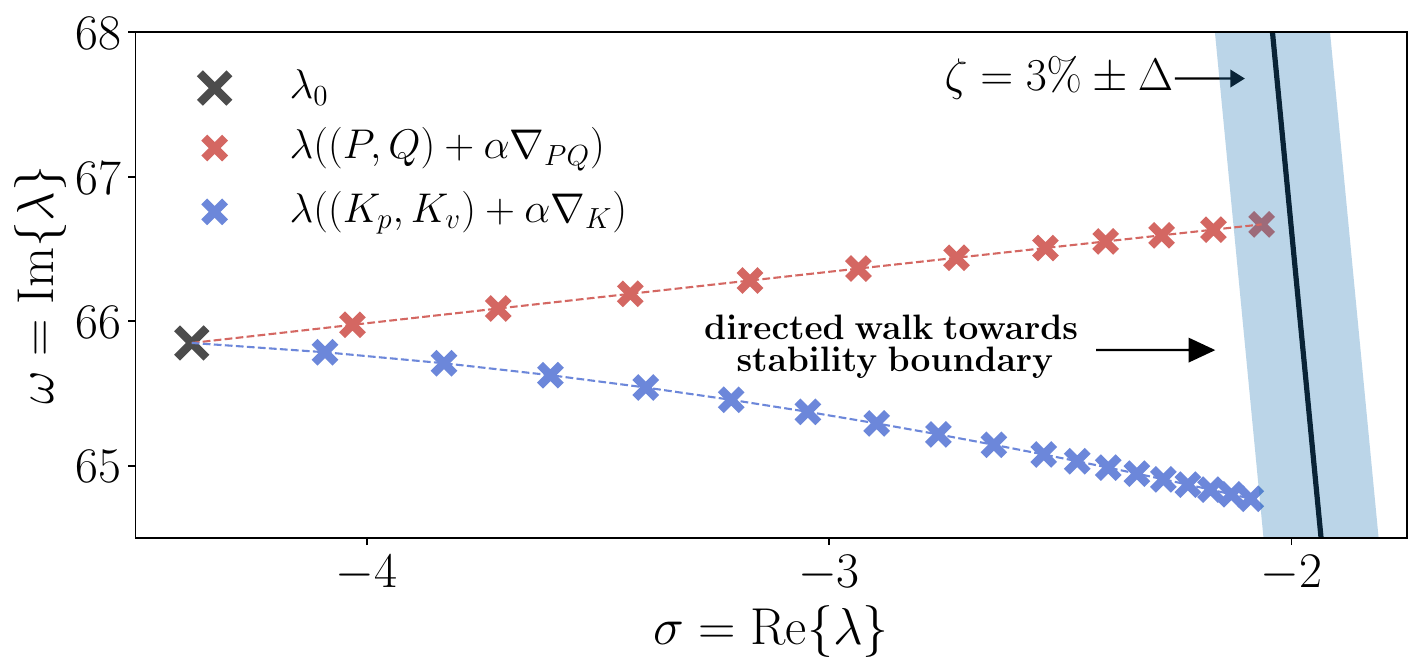}
  \caption{Depicted is a directed walk, starting at a base operating point of $(P_{{\rm ref}},Q_{{\rm ref}},K_{p,f},K_{v})=(1.7,0.1,1.0,0.0)$ and an associated eigenvalue of $\lambda_0=-4.4+j65.8$. Initially, $\zeta_{c,0}=6.63\%$. The top red curve shows the directed walk of the turbine loading set points, while the bottom blue curve shows the directed walk of the voltage and frequency controller gains. Both walks terminate when the eigenvalue damping ratio is within $3\%\pm 0.25\%$.}
  \label{fig:Directed_Walk}
\end{figure}

When dispatching the power set points and control regulation, operators must ensure that local system damping ratios $\zeta_i$
remain above some minimally acceptable threshold (e.g., 3\%). In this paper, we assume this margin has to be maintained across all credible N-1 contingencies (i.e., the loss of any one turbine). The mapping between these parameters and the minimum system damping ratio, however, is intractably dense (see \cref{App_NSWPH}), since it involves solving a series of \mbox{N-1} power flow problems, initializing state variables, linearizing the network, and then computing the damping ratios of this linearized model. To help overcome imposing this computational burden in optimal control problems, we develop a NN mapping between the tunable control parameters and the minimum damping ratio of the NSWPH across all N-1 contingencies:
\begin{subequations}\label{eq: NN_NSWPH}
\begin{align}
{\rm NN}&({\bm x})\rightarrow \hat{\bm{{y}}}_{\bm{x}}\\
\bm{x} &= [P_{{\rm ref}},Q_{\rm ref},K_{p,f},K_{v}]^\top \label{eq:NWSPH_NN_input}\\
 \hat{\bm{{y}}}_{\bm{x}} &=  \hat{\zeta}_{c} \label{eq:NWSPH_NN_output}\\
\underline{\bm{x}} &= [P_{{\rm ref}}^{l}, Q_{{\rm ref}}^{l}, K_{p,f}^{l}, K_{v}^{l}] = [0.0, -0.5, 0.0, 0.0]\\
\bar{\bm{x}} &= [P_{{\rm ref}}^{l}, Q_{{\rm ref}}^{l}, K_{p,f}^{l}, K_{v}^{l}] = [2.0, 0.5, 75.0, 50.0],
\end{align}
\end{subequations}
where $\hat{\zeta_c}$ is the estimated minimum damping across all N-1 contingencies and $\underline{\bm{x}}, \bar{\bm{x}}$ represent the upper and lower bounds of the set points and parameters. In the following subsections, we present the methods we use to sequentially (i) perform data collection and enrichment, (ii) pre-process the data, (iii) regularize the training procedure, (iv) verify model behavior and improve data sampling, and (v) embed the associated model inside an optimization problem.

\subsection{Dataset creation and enrichment}
To collect data for training the NN model of \eqref{eq: NN_NSWPH}, we began by defining a 4-dimensional hypercube 
\begin{align}
\mathcal{H}\!=\![P_{{\rm ref}}^{l},P_{{\rm ref}}^{u}]\!\times\![Q_{{\rm ref}}^{l},Q_{{\rm ref}}^{u}]\!\times\![K_{p,f}^{l},K_{p,f}^{u}]\!\times\![K_{v}^{l},K_{v}^{u}]
\end{align}
where each dimension spans the domains of $P_{\rm ref}$, $Q_{\rm ref}$, $K_{p,f}$ and $K_v$ across their lower ($l$) and upper ($u$) limits. After sampling $\mathcal{H}$ on each dimension a total of $N$ times, we mathematically computed the minimum damping ratio $\zeta_c$ across all N-1 contingencies associated with each operating point $\bm{x}$. We tested three types of sampling techniques:
\begin{itemize}
    \item equally spaced grid (i.e., grid search);
    \item uniform sampling in each dimension;
    \item Latin Hypercube (LHC) sampling.
\end{itemize}
For each routine, we sampled $5^4$, $6^4$, and $7^4$ data points.

While these sampling approaches can evenly sample across the full input domain, we are most interested in collecting data around the pre-defined stability margin. Accordingly, once we computed the minimum damping ratio associated with the $i^{\rm th}$ sampled operating point $\bm{x}$, we performed a directed walk~\cite{Thams2020}. This directed walk iteratively progressed an initial operating point $\bm{x}^{(0)}$ to points $\bm{x}^{(1)}$, $\bm{x}^{(2)}$, etc., each of which mapped to an eigenmode which was incrementally closer to the stability margin $\zeta_c=3\%$. Notably, this process was only implemented if the minimum damping ratio of the initial operating point was sufficiently close to the stability margin (e.g., $3\% \pm 6\%$). The procedure is depicted in \cref{fig:Directed_Walk}, and the details of its implementation are described in~\cref{App_DW}. For each point sampled from the hypercube, and for each terminal directed walk point, we saved both the input $(K_{p,f},K_{v},P_{{\rm ref}},Q_{{\rm ref}})$ and output $(\zeta_c)$ data, but we also saved the gradient information $\nabla_{\bm{x}}=\partial\bm{x}_{i}/\partial\bm{y}_{i}$, since it was also used to effectively regularize the training procedure.

We proceeded similarly for the addition of \textit{network informed} (NI) and \textit{verification informed} (VI) points according to \cref{algo:NI} and \cref{algo:VI}, respectively. For the NI points, we interrupted the training procedure after $1'000$ epochs and sampled a $1'000'000$ possible input data points $\bm{x}'$ in the hypercube. We then evaluated the NN at these points, but retained only predictions within a region close to the stability boundary, defined by the $\hat{\zeta_c} \pm \Delta$, where $\Delta$ determines the adjustable width of the region. We then sampled $N_{\rm samples} = 200$ additional points from this remaining set. These points were then evaluated using the physical model and ultimately added to the training dataset.

For the VI points, we identified regions of consistent classification starting from each of the 16 corner points of the hypercube $\mathcal{H}$. We then sampled a large number of points in the hypercube, and we omitted those input points which fell in the regions that were apriori verified stable or unstable. From the remaining points, we again sampled $N_{\rm samples} = 200$ points which we then evaluated. See \cref{ss:assessment_verification} for the exact implementation of VI sampling.

\subsection{Data pre-processing}
As part of the data handling, we classified the points into five classes for the re-sampling and, later, model assessment. These classes are stable, marginally stable, marginal, marginally unstable and unstable, and they are defined relative to the critical damping ratio as shown in \cref{tbl:test_set_classes}. Alongside the definition, we display the distribution of the classes in the test dataset. We used a very large test dataset ($N = 21^4 = 194'481$) equally spaced in the normalized hypercube. This case may not represent the usual practical setup, as it is rarely feasible to generate such a comprehensive dataset that can be considered as the ground truth. However, we opted for it to show the performance of each model as objectively as possible. Two important notes on this implementation: 1) The test set was only used for ``final" evaluation and was not part of any feedback loop or model selection. 2) The assumption of having the same underlying statistical distribution does not exactly hold for the datasets generated using the LHC and uniform sampling methods. Therefore, this exhaustive test dataset favors the dataset also sampled on a grid; this should be kept in mind when interpreting the final results.

\begin{table}[ht]
  \caption{Size of different stability regions in the test set}
  \label{tbl:test_set_classes}
  \centering
  \begin{tabular}{lccccc}
    \toprule
    Class & Stable & M-Stable & Marginal & M-Unstable & Unstable \\
    \midrule
    $\zeta_c$ & $>\!6\%$ & 6-3.25\% & 3.25-2.75\% & 2.75-0\% & $\leq$0\%\\
    \midrule
    N & 158384 & 9073 & 1399 & 6516 & 19109\\
    \midrule
    Share & 81.4\% & 4.67\% & 0.72\% & 3.35\% & 9.83\%\\
    \bottomrule
  \end{tabular}
\end{table}

For the training/validation split, we opted for an 80/20 split of the datasets, and we fixed how the dataset was split using a constant ``seed" for the utilized random number generator, such that all models performed under identical conditions. The sole data manipulation we conducted in the pre-processing stage was to standardize both the input and the output data. For this, we computed the mean and standard deviation of each input and output in the training dataset, and we applied the same standarization values also to the testing and validation datasets.

\subsection{Model training and gradient-based training regularization}

The objective, or loss, functions of the training problem are defined as
\begin{alignat}{2}
    \mathcal{L}_{y} &= \frac{1}{N} \sum_{j=1}^N || \bm{y}_j - \hat{\bm{y}}_j||_2\label{eq:data_loss}\\
    \mathcal{L}_{J} &= \frac{1}{N} \sum_{j=1}^N || \frac{\partial}{\partial\bm{x}_j}\bm{y}_j - \frac{\partial}{\partial\bm{x}_j}\hat{\bm{y}}_j||_2\label{eq:jacobian_loss}
\end{alignat}
where $N$ stands for the number of data points in the respective dataset. \Cref{eq:jacobian_loss} incorporates the gradient-based regularization as elaborated earlier on.
The associated NN training program is given by
\begin{subequations}
\begin{align}
    \min_{\bm{W}_i, \bm{b}_i} \quad &\mathcal{L}_{y} + \alpha_J \mathcal{L}_{J}\\
    \text{s.t.} \quad &
    \eqref{eq:NN_input}-\eqref{eq:NN_output},
    \eqref{eq:NWSPH_NN_input},\eqref{eq:NWSPH_NN_output}.
\end{align}
\end{subequations}
All models were trained with the Adam optimizer \cite{kingma_adam_2017} with a decaying learning rate $l$ that depends on the epoch $n$ via
\begin{align}
    l(n) = l_0 \times \gamma^n
\end{align}
where $l_0$ represents the initial learning rate and $\gamma$ the learning rate decay. Each NN was trained for in total 3'000 epochs, but, in the spirit of ``early stopping'', the weights and biases that achieved the lowest loss on the validation set were used for model selection purposes. 
For all model forms, we ran a hyper-parameter exploration using the values shown in \cref{tbl:hyperparameter_study}. This was performed by conducting a grid search over the entire parameter space; the best hyper-parameter set was selected by using the validation set. Each combination was run ten times with different weight initialization seeds, and the hyper-parameter set that achieved the best mean of the validation losses on the different initializations was selected. The resulting sets of hyper-parameters are summarized in \cref{tbl:hyperparameter_values} in the appendix. For the final model comparison, each model was run with 100 different initialization using the identified best hyper-parameters. 

\begin{table}[ht]
  \caption{Hyper-parameter options}
  \label{tbl:hyperparameter_study}
  \centering
  \begin{tabular}{lcc}
    \toprule
    Hyper-parameter & Variable & Values\\
    \midrule
    Number of layers & $K$ & [2, 3, 4] \\
    Nodes per layer & $N_k$ &  [16, 32, 64]\\
    Initial learning rate & $l_0$ & [0.005, 0.01, 0.02, 0.05]\\
    Learning rate decay & $\gamma$ & [0.99, 0.995, 0.999, 1]\\
    \midrule
    Jacobian regularization weight & $\alpha_J$ & [0.0, 0.001, 0.01, 0.1, 1.0]\\
    \midrule
    Margin width (for NI, VI) & $\Delta$ & [$\pm 0.25\%$, $\pm 3\%$]\\
    \bottomrule
  \end{tabular}
\end{table}

\subsection{Model assessment and verification formulation}\label{ss:assessment_verification}

For the statistical model assessment, we utilized the previously introduced large test set ($N=21^4$). Besides the total loss across the test set, i.e., \eqref{eq:data_loss}, we also computed the test loss across each of the five classifications $\mathcal{C}_i$ in \cref{tbl:test_set_classes} by substituting $N$ in \eqref{eq:data_loss} only with the $N_i$ points that belong to the respective classification.

\begin{figure}
 \center
 \includegraphics[width=0.9\columnwidth]{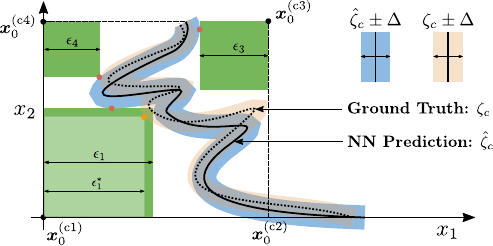}
  \caption{Illustrated is the optimization approach underpinning Verification Informed (VI) sampling in two-dimensional parameter space ($x_1$ and $x_2$). For each of the four corner points (${\bm x}^{({\rm c1})}$ through ${\bm x}^{({\rm c4})}$), the MILP \eqref{eq: MILP_verification} attempts to minimize $\epsilon$ while  the infinity norm (green box) touches $\hat{\zeta}_c\pm\Delta$, which is the marginal stability bound predicted by the NN. The MILP thus generates ``in/stability certificates" in these regions, guaranteeing how the NN will classify all associated internal points. The certificates are then used to filter out future samples. No certificates are generated at ${\bm x}^{({\rm c2})}$, since the stability boundary overlaps closely with the corner point itself. For reference, we also show $\epsilon_1^*$ in the lower box, which is the ground truth value of $\epsilon_1$.}
  \label{fig:Verification}
\end{figure}

The verification problem (for VI sampling) sought to find the smallest distance $\epsilon\!\in\! [0, 1]$ in the normalized hypercube for which the classification threshold (see \cref{tbl:test_set_classes}) is passed. If a given corner point $\bm{x}_0$ was stable, then the threshold was set to $3\% + \Delta$, and if $\bm{x}_0$ was unstable, then it was set to $3\% - \Delta$. For marginal and marginally un/stable points, the verification problem was skipped; \cref{fig:Verification} depicts the procedure in two-dimensional space. The mathematical program associated with a \textit{stable} corner point is
\begin{subequations}\label{eq: MILP_verification}
\begin{alignat}{2}
\min_{\begin{smallmatrix}\hat{\bm{y}}, \breve{\bm{x}}, \bm{z}, \mathbf{b}, \epsilon \end{smallmatrix}} \, \, & \epsilon && \\
    \text{s.t.}  \quad & \hat{\zeta}_c \leq 3\%+\Delta \label{eq: zet_3_margin}\\
    & \left\Vert \breve{\bm{x}}-\breve{\bm{x}}_{0}\right\Vert _{\infty} \leq \epsilon\\
    & \eqref{eq:NN_input}, \eqref{eq:NN_hidden_layers},\eqref{eq:NN_output}, \eqref{Eq:Milp1}-\eqref{Eq:Milp5},
    \eqref{eq:NWSPH_NN_input},\eqref{eq:NWSPH_NN_output},
\end{alignat}    
\end{subequations}
where $\breve{\bm x}$ indicates component-wise normalization ($\breve{{\bm x}}_1={{\bm x}}_1/{{\bm x}}_{1,{\rm max}}$). If the corner point is originally \textit{unstable}, then (\ref{eq: zet_3_margin}) changes to $\hat{\zeta}_c \geq 3\%-\Delta$. As a comparison, we also ran a data-driven version of \eqref{eq: MILP_verification}: instead of using the NN's MILP formulation, we used a sampled collection of points. By approximating $\epsilon$ in that sampled set of $(\bm{x}, \hat{\bm{y}}_{\bm{x}})$ we could compare the effectiveness of the statistical and continuous evaluation techniques. 

\subsection{Embedding problem}\label{ss:embeding_NSWPH}
By employing the MILP reformulation from \eqref{Eq:Milp1}-\eqref{Eq:Milp5}, the learned NN mapping~\eqref{eq: NN_NSWPH} can be directly embedded inside of a variety of otherwise intractable optimization problems. One such problem which motivates this work is the co-optimization of droop parameters in the context of a regional planning study (e.g., as in~\cite{Tosatto:2021}). Such planning studies often necessarily utilize overly simplistic linear models of frequency dynamics. Through NN embedding, however, linear models can be replaced by \textit{piecewise} linear NN-based models which have the capacity to be far more accurate~\cite{Kody:2021}.


In the NSWPH case, for any targeted nominal power injection $P^{\star}_{\rm ref}$, $Q^{\star}_{\rm ref}$ from the wind turbines, we can find regions of droop parameters around a given parameter setting $K_{p,f,0},K_{v, 0}$ which are predicted to yield definitely-stable system behavior: 
\begin{subequations}\label{eq:max_droop}
\begin{alignat}{2}
\min_{\begin{smallmatrix}\hat{\bm{y}}_{\bm{x}}, \breve{\bm{x}}, \bm{z}, \mathbf{b}, \epsilon \end{smallmatrix}} \, \, & \epsilon && \\
{\rm s.t.}\quad\; & \bm{x}=[P_{{\rm ref}}^{{\star}},Q^{{\star}}_{{\rm ref}},K_{p,f},K_{v}]\\
& \bm{x}_0 =[P_{{\rm ref}}^{{\star}},Q^{{\star}}_{{\rm ref}},K_{p,f,0},K_{v,0}]\\
& \hat{\bm{y}}_{\bm{x}}=\hat{\zeta}_{c}\leq 3\% + \Delta\\
& \left\Vert \breve{\bm{x}}-\breve{\bm{x}}_{0}\right\Vert _{\infty} \leq \epsilon\\
& \eqref{eq:NN_input}, \eqref{eq:NN_hidden_layers},\eqref{eq:NN_output}, \eqref{Eq:Milp1}-\eqref{Eq:Milp5}.
\end{alignat}
\end{subequations}

Alternatively, for fixed droop $K^{\star}_{p,f}$, $K^{\star}_{v}$, we can ensure regions of power injections around a given power injection $P_{\rm{ref},0}$, $Q_{\rm{ref},0}$ yield satisfactory damping ratios:

\begin{subequations}\label{eq: min_damp}
\begin{alignat}{2}
\min_{\begin{smallmatrix}\hat{\bm{y}}_{\bm{x}}, \breve{\bm{x}}, \bm{z}, \mathbf{b}, \epsilon \end{smallmatrix}} \, \, & \epsilon && \\
{\rm s.t.}\quad\; & \bm{x}=[P_{{\rm ref}},Q_{{\rm ref}},K_{p,f}^{{\star}},K_{v}^{{\star}}]\\
& \bm{x}_0 =[P_{\rm{ref},0}, Q_{\rm{ref},0},K_{p,f}^{{\star}},K_{v}^{{\star}}]\\
& \hat{\bm{y}}_{\bm{x}}=\hat{\zeta}_{c}\leq 3\% + \Delta\\
& \left\Vert \breve{\bm{x}}-\breve{\bm{x}}_{0}\right\Vert _{\infty} \leq \epsilon\\
& \eqref{eq:NN_input}, \eqref{eq:NN_hidden_layers},\eqref{eq:NN_output}, \eqref{Eq:Milp1}-\eqref{Eq:Milp5}.
\end{alignat}
\end{subequations}
The solutions of these embedding problems are illustrated in the subplots of~\cref{fig:embedding_problems}.

\subsection{Notes on the implementation}
The entire framework, which is provided on our Github~\cite{Stiasny_github:2022}, is implemented in Python 3.8 using standard libraries. All parts related to the training of the model utilize PyTorch \cite{pytorch:2019}. The optimization problems are formulated in Pyomo \cite{hart2011pyomo, bynum2021pyomo} and solved using Gurobi \cite{gurobi}. The logging and tracking of all training runs, in particular for the hyper-parameter tuning, is conducted with WandB~\cite{wandb}. The training and optimization were performed on the machines in the high performance computing (HPC) cluster at the Technical University of Denmark (DTU) \cite{DTU_DCC_resource} that are configured as nodes with 2xIntel Xeon Processor 2650v4 (12 core, 2.20GHz) and 256 GB memory.








\section{Results}\label{sec:results}

The following paragraphs show the previously elaborated steps in greater detail by illustrating important aspects of the procedure. It is important to realize that we encountered competing objectives when training the NN: on one hand, we aimed to achieve an accurate predictor across the full input domain in order to use the NN as a representative function in an eventual embedding problem. On the other hand, we tried to capture the region around the critical damping ratio as closely as possible. With these competing objective in mind, we used data associated with the full input space as the decision criterion for all model selection decisions, but by design, the methods that sampled \textit{additional} training points targeted data close to the stability boundary.


\subsection{Problem characteristics - class imbalance}

\begin{figure}[ht]
    \centering
    \includegraphics[width=1.0\linewidth]{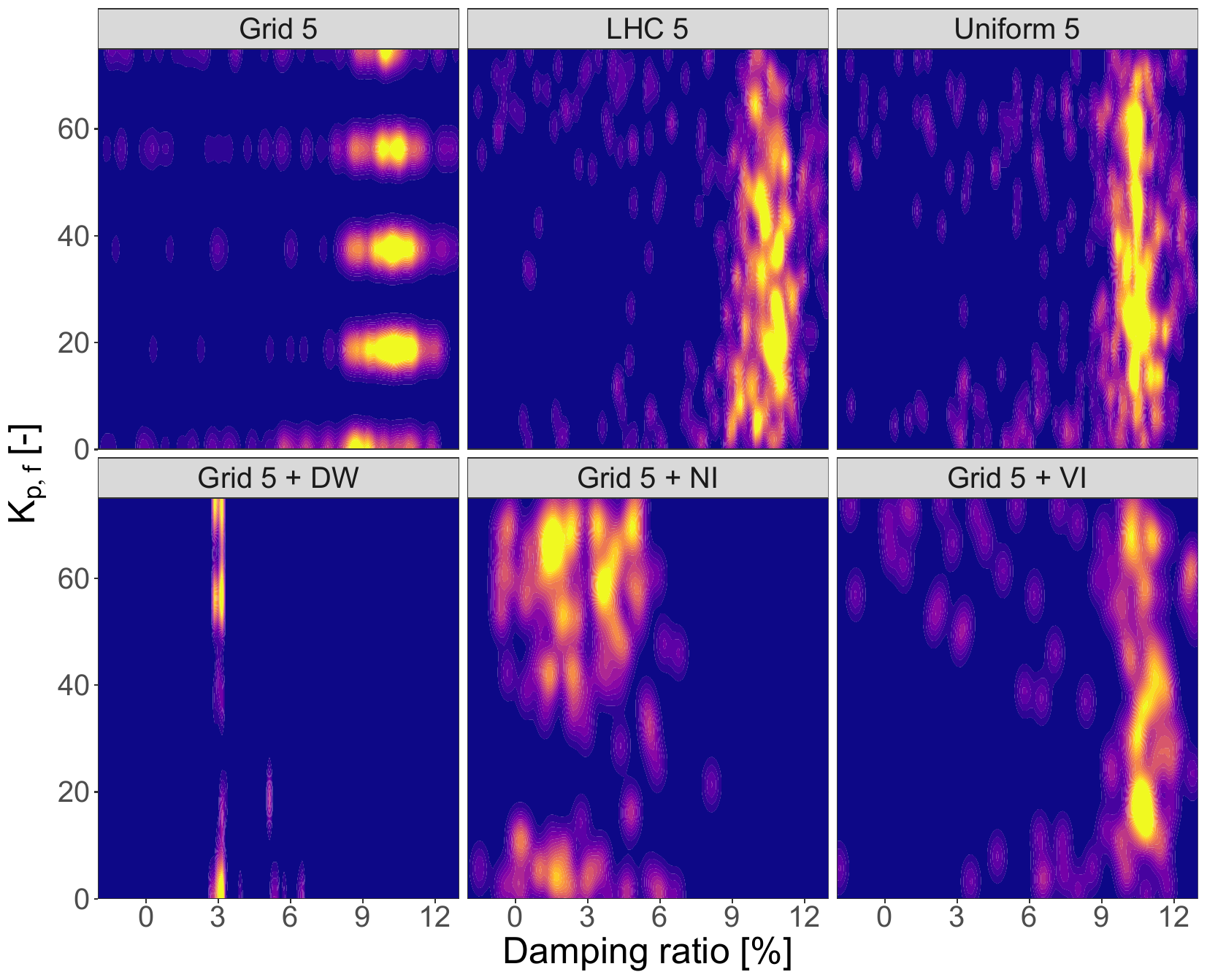}
    \caption{Visualization of the sampled data distributions for six different datasets. The bright yellow areas signify high densities of sampled points whereas the dark blue areas contain no data samples. The top three distributions were generated using conventional sampling techniques (i.e., Grid sampling, Latin Hypercube sampling, and Uniform sampling). The bottom three distributions  were generated using directed walk (DW), Network Informed (NI) and Verification Informed (VI) sampling.}
  \label{fig:variable_distribution}
\end{figure}

Before diving into the direct comparison of the methods, we take a brief look at the characteristics of the problem, namely the mapping $\bm{f}$ from the input to the output domain. The fundamental problem in the dataset creation becomes apparent in \cref{fig:variable_distribution}, which illustrates the relationship between the input parameter $K_{p,f}$, the damping ratio $\zeta$, and the number of data points collected in each region. With simple sampling methods (which are used to generate the upper row in \cref{fig:variable_distribution}), the vast majority of data points lie well within the \textit{stable} region; this also means that very few data points fall in or close to the marginal region. In contrast, a very high percentage of the DW dataset (lower left) hits the marginal stability region, as it was designed to do. Similarly, the NI resampling ($\Delta = 3\%$) creates a much more targeted sample distribution. Lastly, the VI resampling ($\Delta = 0.25\%$) is not as successful in creating a more balanced distribution, mainly due to the fact that it only can omit about 40\% of the input domain. \Cref{tbl:dataset_classes} in the Appendix reports a more detailed overview of this class imbalance, and it shows how the resampling distribution varies across different training runs associated with the NI and VI approaches.

Understanding these problem specific dataset characteristics is a crucial step for the subsequent performance assessment and analysis. This also highlights why the data pre-processing step should also include thorough data exploration; simple tools, such as the regional classification of \cref{fig:variable_distribution} or correlation analysis, can already provide vital insights. 


\subsection{Comparison of the model performance}

\begin{figure*}[ht]
    \centering
    \includegraphics[width=1.0\linewidth]{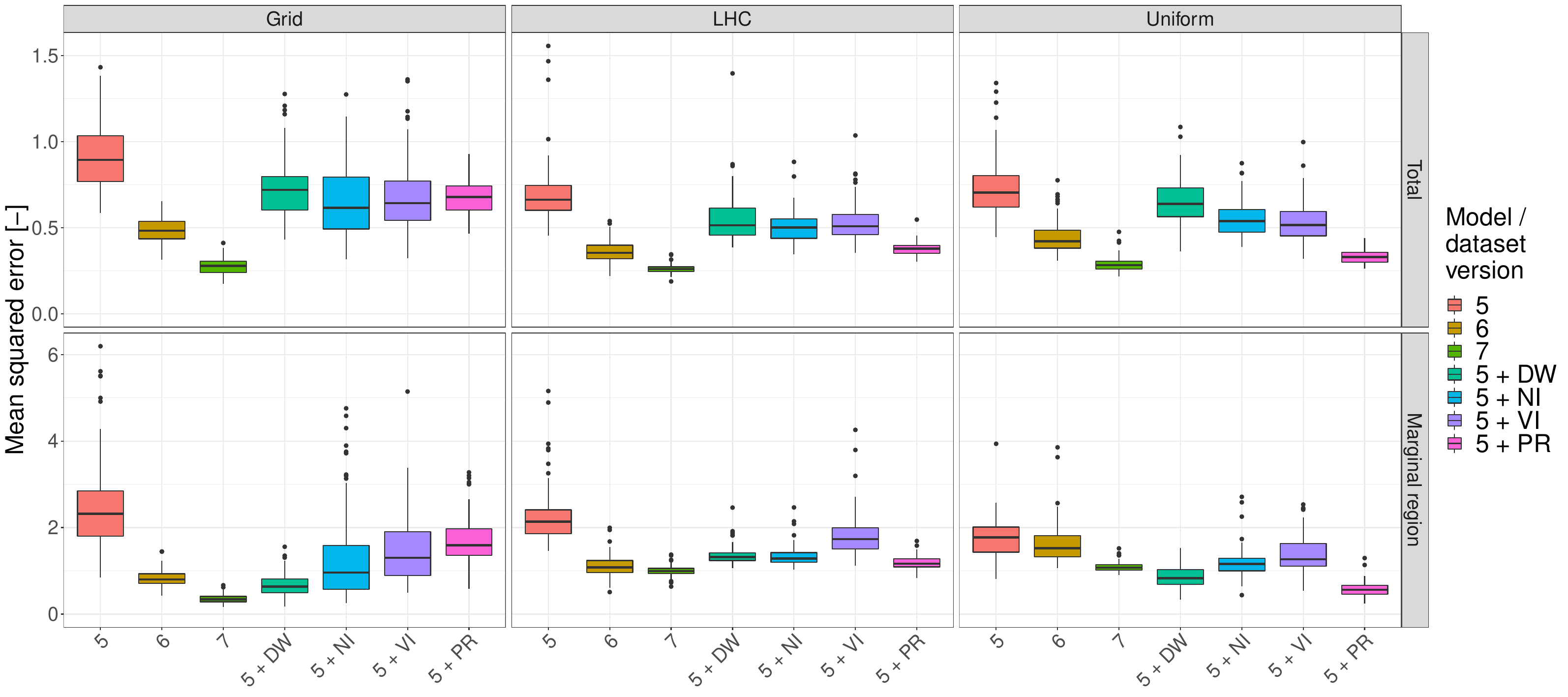}
    \caption{Statistical model assessment of the selected models listed by sampling method and dataset/model version. The boxplots represent 100 runs with different weight initializations. Assessment by the mean squared error on the entire test set (total) and on the subset in the marginal region. The x-axis labels of ``5", ``6", and ``7" refer to the number of data points sampled: $5^4$, $6^4$, and $7^4$, respectively.
    \\The following seven outliers (total MSE, MSE in marginal region) are not displayed for visualization reasons:\\Grid 5: (3.5, 15.1) - LHC 6: (3.1, 18.1) - LHC 5 + VI: (3.2, 19.4), (3.3, 19.5), (3.1, 19.9) - Uniform 5: (3.4, 17.9) - Uniform 5 + VI: (3.1, 18.4)}
    \label{fig:error_comparison}
\end{figure*}

With these considerations, we move to the central results of this work: the model assessment in order to compare the effects of the previously presented approaches. \Cref{fig:error_comparison} visualizes the outcome of the statistical performance assessment of the different variations of the models for the two primary metrics of interest: the overall mean squared error (loss), and the loss in the marginal stability regions that we target to improve. The Box-Whisker plots stem from NN training repetitions across different initializations, subsequently called ``runs", of the NN's weights and biases. Across the different sampled base datasets, it becomes clear that simply adding more data points improves the test accuracy (the loss on the test set) in terms of the mean, but also for the standard deviation, across the different runs. This is simply the effect of better model generalization and less dependence on a good training run by virtue of a ``lucky" initialization. The addition of the physics regularization (PR) shows a similar effect for the datasets stemming from LHC and uniform sampling. This is a clear sign for an effective regularizer -- in contrast, the PR for the Grid dataset is much less powerful. This specific effectiveness, which depends on the sample distribution, may stem from the following consideration: the gradient regularization helps the NN interpolate better in between data points. However, for the Grid dataset, the data points are already implicitly carrying this gradient information, albeit not very accurately. Hence, the interpolation to points that lie not on the lattice is not very much aided by the additional regularizer, i.e., the additional information content is not as high as for the other sampling strategies.

We have to note that these results are not sufficient to draw conclusions about the best sampling approach, as only a single dataset was created for the LHC and Uniform approach. For conclusions on the sampling efficiency, these tests would need to be repeated several times with different realizations of the sampled datasets. Nonetheless, it is striking that both the NI and VI approach yield more consistent results, i.e., smaller standard deviation, for LHC and Uniform sampling than for the grid dataset. The cause probably lies in the much less accurate predictions of some training runs at the time when the additional points are sampled. Hence, the additional points will not be as close to the target region, which leads to lower performance improvement. In this regard, we can also observe that the NI approach has an advantage over the VI approach across all sampling strategies when measured by the test loss in the marginal region. This is not surprising, as we showed in \cref{fig:variable_distribution}: the NI approach is much more effective at sampling in the marginal region. The VI approach can reduce the size of the input domain, but mostly only by a 40-50\% measure in terms of the volume of the unit hypercube. \Cref{tbl:dataset_classes} in the Appendix supports this observation, by analyzing the share of the (additional) samples of each class. Unsurprisingly, the DWs yield the best performance among the ``resampling" strategies when considering the loss in the marginal region. This is particularly noticeable for the Grid dataset, as the DW outperforms both the NI and VI approaches.

Ultimately, these analyses show that by introducing additional elements into the ML framework, in particular feedback loops, we can create well performing models without the need for greatly increased sampling efforts. To judge the efficiency of these methods, we consider how well the simple sampling strategies perform given an increased number of points (approximately double and four times for datasets with $6^4$ and $7^4$ sampled points). Whereas the performance improvements for DW, NI, and VI on the overall test loss can be attributed, to some extent, to the additional number of data points (usually around 200), the test loss in the marginal region clearly shows that a target sampling of these ca. 200 points can lead to performances that would need significantly more additional data points with conventional sampling methods. Furthermore, if an effective regularizer can be provided, such performance boosts can be obtained without additional data. \Cref{tbl:error_comparison} in the Appendix presents all the results in a numerical form.


\subsection{Hyper-parameter tuning and model selection}
\begin{figure}
    \centering
    \includegraphics[width=\linewidth]{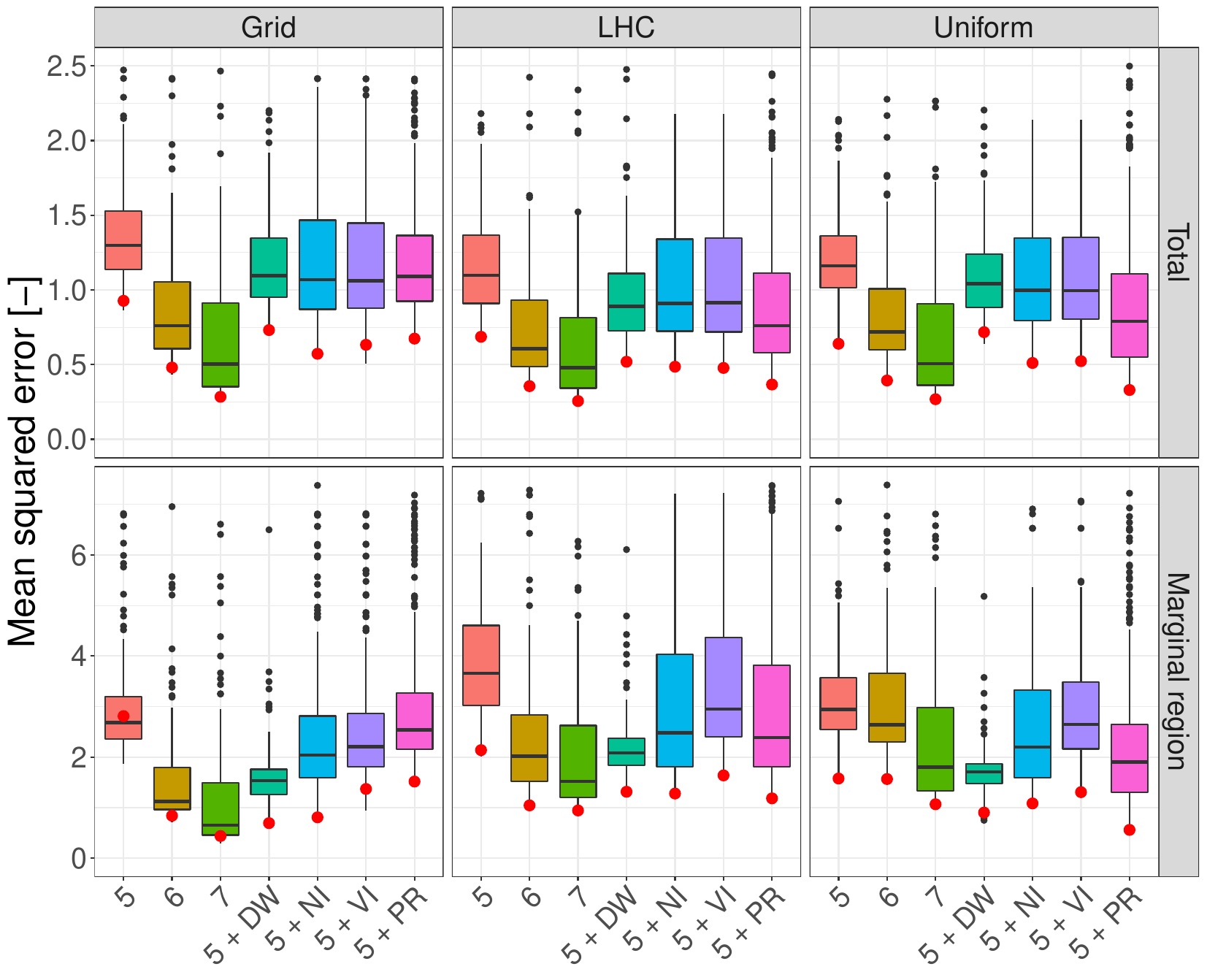}
    \caption{Quality of the hyper-parameter tuning. The boxplots represent the average performance of each hyper-parameter combination evaluated on the test set (see \cref{tbl:hyperparameter_study}). The red dot shows the selected model based on the validation loss of the particular dataset (see \cref{tbl:hyperparameter_values}). The fact that the selected hyper-parameters do not always achieve the best performance when evaluated on the test set underlines the importance of the dataset splitting. Same coloring as in \cref{fig:error_comparison}.}
    \label{fig:hyperparameter_quality}
\end{figure}

The previously presented results represent only the model assessment step; the model learning and, in particular, the model selection steps are often given much less attention. Therefore, we want to stress the importance of a thorough hyper-parameter tuning and model selection procedure. The presented framework and its implementation allow for easily running a large number of experiments with the sole restriction being the computational burden. The HPC cluster we used at DTU allowed us to systematically test all combinations of hyper-parameters listed in \cref{tbl:hyperparameter_study}. After running these combinations with 10 different initialization seeds each, we used the mean of the loss across the respective validation set as the model selection criterion, which led to the selection as presented in \cref{tbl:hyperparameter_values} in the Appendix. At this point, the aforementioned importance of a distributionally invariant split between training, validation, and testing data becomes relevant. For the model to generalize well, model selection must not happen based on the test set that is used for the performance assessment -- otherwise, the assessment will be biased. Having distributionally invariant subsets of the datasets will allow us to choose well performing models nonetheless. In \cref{fig:hyperparameter_quality}, we show what the performance assessment distributions that the different hyper-parameter combinations would have yielded (boxplots), together with the hyper-parameter setting that was chosen based on the above mentioned model selection criterion (red dot), i.e., the loss on the validation set. We can observe that in all cases the overall test loss is close to the best possible model, and similarly for the test loss in the marginal region. However, there are a few instances where there could have been a better hyper-parameter selection, e.g., for the Grid VI model. This insight, however, is only an analysis in hindsight, and it must not influence the model selection process. On the other hand, we can also see how a poor hyper-parameter selection could have led to much worse model performance and hence completely cancel out the performance benefits of the various methods proposed in this paper and invalidate all previously presented results. 

It should be noted that the extent of this hyper-parameter study is usually beyond the scope of a practical implementation. However, for a reliable performance comparison, in particular if conclusions about the superiority of a method are to be drawn, a sufficient exploration of the utilized hyper-parameters and a rigorous model selection procedure must be performed. Working in a fixed framework, as presented, can greatly reduce the effort to do so and at the same time place safeguards against (even unconsciously) biasing the results.

\subsection{Statistical vs. continuous model assessment}

A central element in the framework is the reformulation of the NN as a MILP. We use it for the VI resampling, and for the embedding problem, and it can be used also in performance assessment. To show the benefit that we can draw from the reformulation, we look at how a statistical evaluation (i.e., simply passing data points through the NN to assess stability regions) compares to the MILP verification (i.e., using the MILP reformulation to rigorously assess stability regions).

\begin{figure}[!th]
\centering
\subfloat[Stable corner point $\bm{x}_1$]{\includegraphics[width=\linewidth]{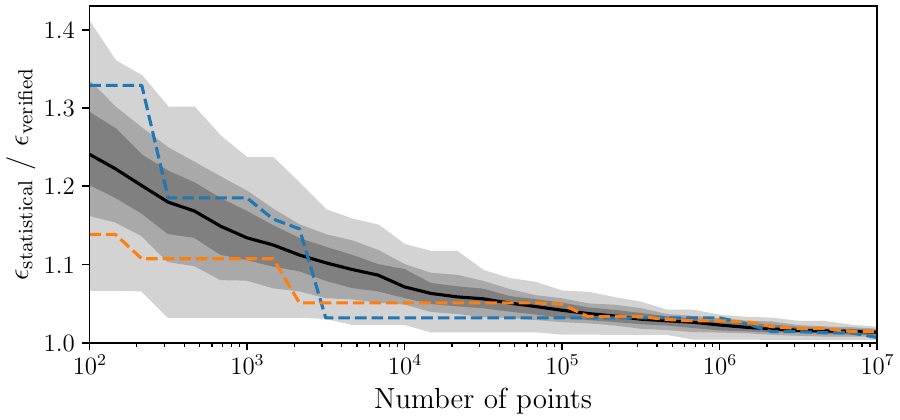}%
\label{fig:verification_corner_0}}
\hfill
\subfloat[Unstable corner point $\bm{x}_2$]{\includegraphics[width=\linewidth]{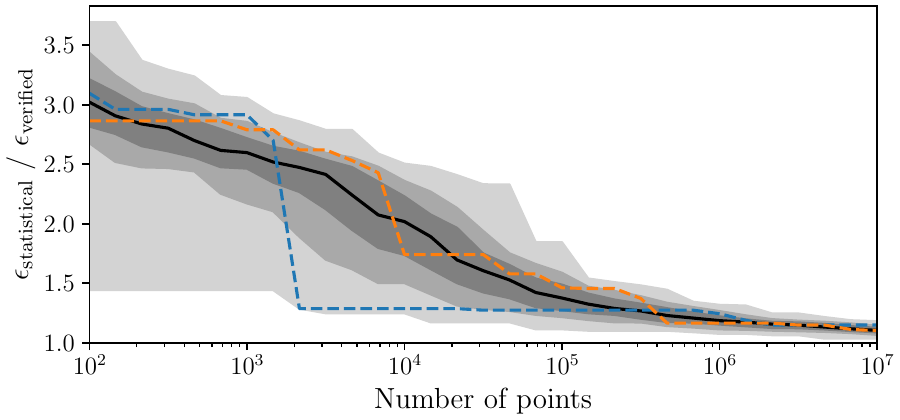}%
\label{fig:verification_corner_1}}
\caption{Comparison of the statistical estimation of $\epsilon$ value, i.e., $\epsilon_{\rm{statistical}}$, to the value stemming from the verification, i.e., $\epsilon_{\rm{verified}}$, (\ref{eq: MILP_verification}), for a stable (a) and an unstable (b) region of operation. This ratio is shown across an increasing number of points used in statistical evaluation. The dashed lines show two example evaluations, while the solid dark lines shows the median across 100 different samplings for $\epsilon_{\rm{statistical}}$. The shaded gray areas represent the range between the [0th, 100th], [10th, 90th], and [25th, 75th] percentiles.}
\label{fig:statistical_vs_verification}
\end{figure}

As an example, we look at the calculation of a certified region starting from two corner points, namely $\bm{x}_1 = [0.0, -0.5, 0.0, 0.0]$ and $\bm{x}_2 = [0.0, -0.5, 0.0, 50.0]$; the former should result in a stable region and the latter in an unstable one. We run the verification according to \eqref{eq: MILP_verification} and subsequently compare the result to the size of the region that a statistical evaluation would yield for increasing the number of tested points. \Cref{fig:statistical_vs_verification} shows the results, here plotted as the distribution across 100 different samplings. It is intuitive that the gap between the verified behavior ($\epsilon_{\rm{verified}}$) and its empirical estimation ($\epsilon_{\rm{statistical}}$) closes with the increase in the number of sampled points. However, the plot should also make clear that the empirical value can change drastically with the addition of only a single point. It so happened for the blue curve in the lower panel of \cref{fig:statistical_vs_verification}: around the 1100th point, the gap abruptly diminished to then stay constant until the millionth point. In contrast, the gap for the orange sample only reaches this level after many hundreds of thousands of points are used. Furthermore, a good estimation for one gap does not translate into a good estimation for any other gap, as we can observe when comparing the ``good" estimation of the orange sample in \cref{fig:verification_corner_0} with the ``bad" one in \cref{fig:verification_corner_1}.

This example illustrates that relying on empirical evaluations of the behavior of neural networks requires a large number of points to become statistically trustworthy. And even then, there is still a chance for very small regions that yield a contradicting prediction -- only by verifying a NN with optimization tools can we be certain about its definite behavior. In case such a contradicting prediction occurs, we can then furthermore compare it to the ground truth and identify it either as an adversarial example or confirm the seemingly contradicting prediction. 

\subsection{Embedding possibilities}

\begin{figure*}[!th]
\centering
\subfloat[$P_{{\rm ref}}^{{\star}} = 1.2\, \rm{p.u.}$, $Q_{{\rm ref}}^{{\star}} = -0.5\, \rm{p.u.}$ ]{\includegraphics[width=0.34\linewidth]{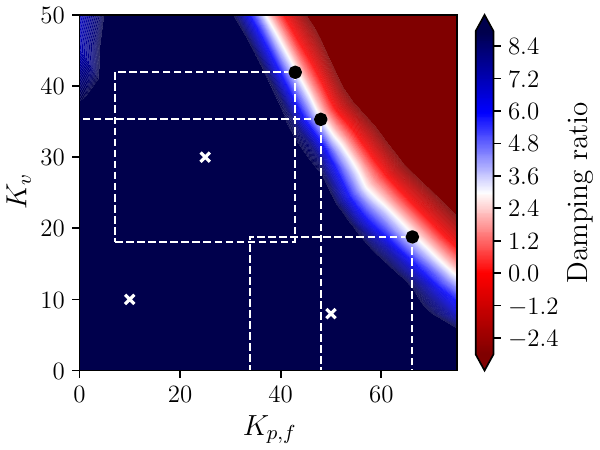}%
\label{fig:embedding_problems_power_low}}
\hspace{1cm}
\subfloat[$P_{{\rm ref}}^{{\star}} = 2.0\, \rm{p.u.}$, $Q_{{\rm ref}}^{{\star}} = 0.5\, \rm{p.u.}$]{\includegraphics[width=0.34\linewidth]{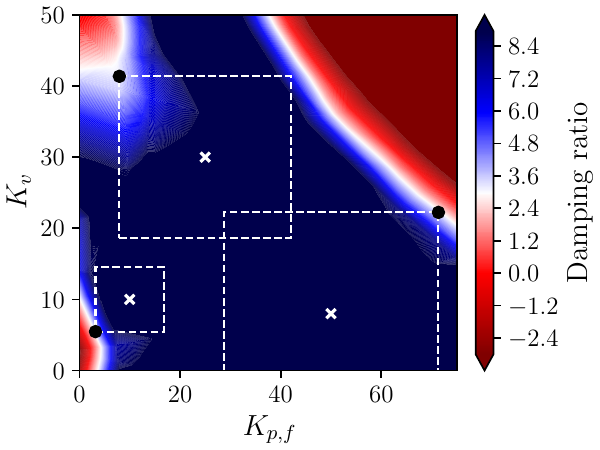}%
\label{fig:embedding_problems_power_high}}
\hfill
\subfloat[$K^{\star}_{p,f} = 0.0$, $K^{\star}_{v} = 4.0$]{\includegraphics[width=0.34\linewidth]{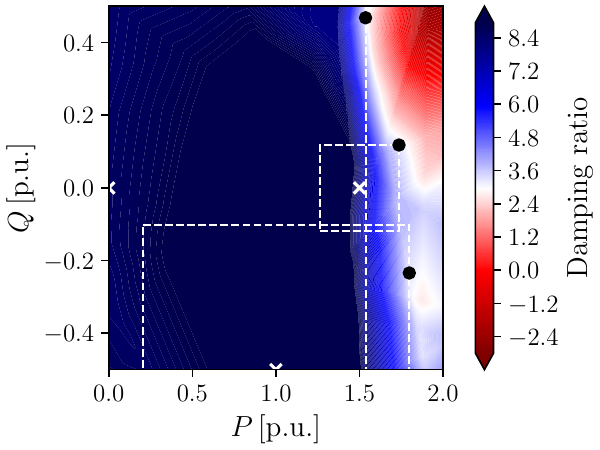}%
\label{fig:embedding_problems_control_low}}
\hspace{1cm}
\subfloat[$K^{\star}_{p,f} = 60.0$, $K^{\star}_{v} = 40.0$]{\includegraphics[width=0.34\linewidth]{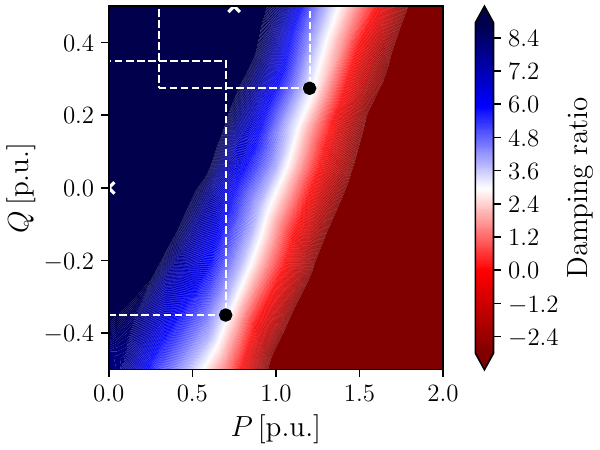}%
\label{fig:embedding_problems_control_high}}
\caption{Verified regions (dashed white lines) for the various embedding problems. Blue regions are operationally acceptable, while red regions are unacceptable. The black dots signal the points of changing classification.}
\label{fig:embedding_problems}
\end{figure*}

As a last topic of the results we come back to the motivation of this work: training a NN as a surrogate function which we then can embed into other problem such as described in \cref{ss:embeding_NSWPH}. \Cref{fig:embedding_problems} presents two illustrative results for each of the posed problems. \Cref{fig:embedding_problems_power_low,,fig:embedding_problems_power_high} show the verified regions of control parameters that lead to stable behavior for a fixed power injection $P^{\star}_{\rm ref}$, $Q^{\star}_{\rm ref}$. In the inverted setting, where droop parameters $K^{\star}_{p,f}$, $K^{\star}_{v}$ are fixed, \cref{fig:embedding_problems_control_low,,fig:embedding_problems_control_high} identify admissible power injections which could be included, for example, in an OPF setting.

Following up on the previous section of statistical versus verified behavior, the shown regions outlined by the dashed white lines are verified, whereas the contours in the plot stem from NN predictions evaluated on an extremely fine grid (akin to a form of statistical evaluation). Either of the options could be used; however, as the input problem's dimensionality increases, the latter becomes harder and harder to realize.







\section{Discussion}\label{sec:discussion}
We view the proposed framework, along with the considerations presented throughout this paper, as important steps towards bringing Power Systems and Machine Learning (ML) closer together and enabling the use of ML in safety-critical applications. Borrowing from the ML community, one could find plenty of generic workflows and implemented frameworks and \cite{paleyes_challenges_2022} reviews a number of case studies -- however, for a successful adoption of ML methods to power systems, we need to tailor these approaches to our specific needs. At the same time, the sheer number of specific problems in the power systems domain makes generic solutions, as well as community-wide benchmarks, difficult to implement, especially since the primary goal of using ML in power systems is still driven by specific applications.
We should, nonetheless, strive towards clarity, comparability, and transparency in our efforts in ML, because
the alternative (e.g., unclear assumptions, improper dataset handling, or badly tuned algorithms) will certainly not strengthen efforts to adopt ML in safety-critical applications. To establish such practices, we believe that three pillars will be crucial: i) code, ii) data, and iii) communication. In all of these aspects, a framework, such as the one proposed in this paper, can be helpful, and the following paragraphs will briefly discuss how.

i) \textit{Code}: First and foremost, when it comes to publishable research, the produced code should be made publicly available; this not only helps demonstrate the strengths of the proposed algorithm (which is very difficult to prove otherwise), but it can also serves as a benchmark for other researchers to compare with. Second, it should not be underestimated how much a framework can help structure the code and thereby make it more adaptable. Ideally, each of modules shown in \cref{fig:workflow} is treated as a stand-alone piece of code. This not only allows one to interchange methods more easily, e.g., the sampling method for the dataset creation, but it also allows one to be more aware of the information flow between modules. This means that each of the arrows in \cref{fig:workflow} is associated with passing on some information; this can range from datasets to parameters to performance assessments. By being more conscious about which information is being passed on and by trying to reduce these exchanges to the necessary minimum, we become less prone to introducing biases in the process and it gets easier to check for such undesired information flows. The classic example is that the test dataset should not enter the model learning but only the model assessment. And especially if we introduce more loops in the framework, it must be ensured that, e.g., the VI or NI approach do not make use of the test dataset. A side-effect of modularizing code is that the workflow becomes more and more generic which immensely simplifies conducting hyper-parameter studies and the comparison of methods. In the provided code base \cite{Stiasny_github:2022}, a small configuration file is used to setup and run all the different setups that are reported. Thereby, the risk of errors greatly reduces and the repeatability of experiments in case of changes in a module is significantly simplified.

ii) \textit{Datasets}: It should be apparent by now that the used datasets are of utmost importance. Providing access to the models that have been used for the dataset creation (or at least, the resulting datasets) will become a necessity. In case proprietary reasons prevent a release, mock datasets that resemble the used datasets should be tested and provided. Furthermore, it is advised to use so-called seeds in all processes associated with random numbers, e.g., for the train/validation/test split. It ensures the reproducability of the results, and it reminds us to consider how general the drawn conclusions can be. In the presented results, the LHC and uniform sampling approaches have only been performed once, as tests with multiple sample realizations would have increased the computational burden even further. This design choice, which can be challenged, needs to be accounted for when comparing the results across the different sampling methods. A conclusion that would claim the superiority of, e.g., uniform over LHC sampling, is simply not trustworthy based on our results, as this conclusion might have been an artifact of the randomly chosen datasets.

iii) \textit{Communication}: One of the biggest challenges is to efficiently communicate at the intersection of ML and power systems. It begins with terminology: some of the terms are community specific, some overlap, and some are over-loaded, meaning that the same word can take multiple meanings. A simple example is the word `model': we tried to avoid confusion by using NSWPH model or physical model for the power system model and ML model for the learned neural network, but still it is not always easy to be clear. Furthermore, with all additional terminology, there is a degree of ambiguity and it will take time to agree upon how the power systems community will use new terms introduced from the machine learning jargon. For such a setting, a framework as proposed can be helpful as a starting point. Beyond simple terminology, building trustworthy ML also requires a common understanding of what needs to be presented in a ML related research article. By requiring, e.g., an overview of the hyper-parameter tuning or basic statistical properties of the dataset, we might be able to establish good practices and raise awareness of the associated issues among the entire community.    

In terms of technical aspects of the presented work, we want to briefly discuss two aspects that have not been addressed in the Results Sections: The overall computational effort and implications for practical work.

On the aspect of the computational effort, it is important to distinguish between the entire workflow and individual steps involved. From a high-level perspective, there is a trade-off to be made between model performance and computational effort. The simplest example of this is for instance that with the possibility to generate additional data points, it should be possible to create a better performing model, but at the cost of computing the additional data points. Therefore, an honest comparison of methods pays tribute to this trade-off. This can, for example, be achieved by reporting the sensitivity of a method to the dataset size, i.e., by training with varied dataset sizes, as we did. If a method, such as the NI approach, can reach a similar performance improvement as a simple doubling of the dataset size would while only requiring a third of the necessary additional simulations, we can consider the method a true improvement. This comparison was rather straightforward as it did not involve measuring how long it takes to simulate a data point. As soon as we enter comparisons between methods that involve computationally intense tasks, the actual implementation becomes a source of uncertainty that is hard to control. For example, the dataset creation can be sped up significantly by performing all calculations in C-compiled code. On the other hand, the VI approach is strongly dependent on the NN size but also its sparsity. The resulting assessment of whether a larger dataset or the VI approach is computationally more efficient becomes arbitrary as long as these comparisons are not performed with open-source state-of-the-art implementations. To alleviate this complication, one can instead revert to qualitatively analyzing the share of each module or step in the process to identify the biggest burdens and how the proposed methods affect these. In the presented study, the dataset creation clearly dominates the burden of the training (also including the physics-regularisation). For the verification, the picture is mixed, as it quickly requires the biggest computational effort for the unfavorable cases of large NNs. But as mentioned previously, with an improved setup, i.e., a sparser NN, this relation could easily be proven wrong. Accordingly, we did not strive for minimal computation time, and, hence, did not place emphasis on analyzing it in detail. Another, more formal alternative can be to focus on analyzing algorithms in terms of their algorithmic complexity such that their scalability can be judged.

For a practical implementation of the framework, the extent of the hyper-parameter study can be significantly reduced. Instead of a full grid search, which we conducted for a complete picture, it is much more practical to first run a few tests with random hyper-parameters drawn from a larger range before moving to a reduced range with a more thorough grid search. In the study, the low learning rates and large learning rate decays could have for example been discarded early on. A slightly advanced method is to employ Bayesian hyper-parameter tuning. These comments, though, shall not encourage one to take short-cuts on hyper-parameter tuning, but instead conduct it in an efficient manner. A final note on how the suggested methods might be used: From a methodological point of view, the physics-regularization and the NI resampling approach appear to be the most promising at the moment. The former appeals due to its simplicity and avoidance of the (expensive) creation of additional data points, the restriction being that the required values, here derivatives, are available. On the other hand, the NI resampling approach stands emblematic for the idea of ``closing the loop'' as it combines the cheap evaluation and selection of candidates via the NN and a targeted dataset enrichment strategy. While not as targeted as the directed walks, it achieves a similar effect without the majority of its cost. 
Regarding the VI approach, improving the size of the rejected areas, e.g., by including points other than the corner points, will be crucial to use it for general resampling purposes. The true potential of verification-based approaches lies in the structured search for adversarial examples and the possibility to provide regions of verified behavior for the embedding problems; these are elements that a purely statistical assessment cannot provide. 

\section{Conclusion}\label{sec:conclusion}
In this paper, we identified key challenges associated with building high-fidelity ML models for power system applications. We then integrated state-of-the-art solutions to these various challenges into a single coherent framework. The proposed workflow at the heart of this framework utilizes tools such as physics regularization, hyper-parameter sweeping and loop-closing in order to efficiently generate ML models which are both trustworthy and generalizable. We tested the proposed framework on a realistic model of the North Sea Wind Power Hub. Through extensive analysis, we quantitatively demonstrated the benefits that can be provided by the various elements of the proposed framework. We also illustrated how derived ML models can be used to generate useful embedding (e.g., \eqref{eq:max_droop}) and interpretable mappings (e.g., \cref{fig:embedding_problems}) for power system operators. Finally, we concluded with discussions concerning both the importance of proper ML practices within the power systems community and avenues for future progress.

\appendix
\subsection{North Sea Wind Power Hub: Technical Details}\label{App_NSWPH}
\begin{figure}
 \center
 \includegraphics[width=0.8\columnwidth]{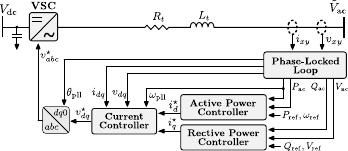}
  \caption{Shown are the control loops associated with a Voltage Source Converter (VSC) interconnecting AC and DC terminals. The details of the active and reactive power control loops are depicted in Fig. \ref{fig:NSWPH_Control}.}
  \label{fig:VSC_Control}
\end{figure}

\begin{figure}
 \center
 \includegraphics[width=0.8\columnwidth]{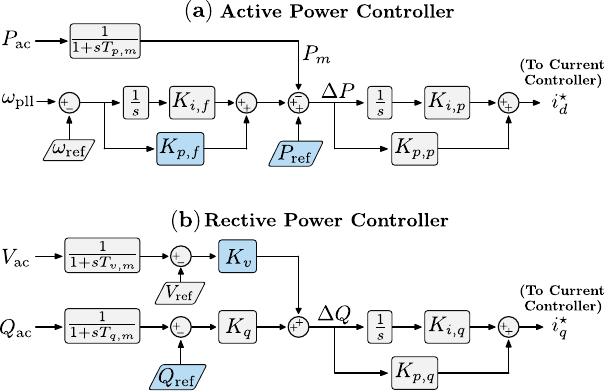}
  \caption{Panel (\textbf{a}) depicts the active power control loop, where $K_{p,f}$ controls the proportional active power injection response to frequency deviations, as measured by a phase-locked loop (PLL), while $P_{\rm ref}$ is the adjustable active power reference. Panel (\textbf{b}) depicts the reactive power control loop, where $K_{v}$ controls the proportional reactive power injection response to voltage deviations, while $Q_{\rm ref}$ is the adjustable reactive power reference.}
  \label{fig:NSWPH_Control}
\end{figure}

Within the next decade, governments and grid operators around the North Sea plan to install offshore wind power hubs with never-before-seen capacities. The study on the ``North Sea Wind Power Hub" (NSWPH) in~\cite{Misyris:2020_NSWPH} details one possible configuration of this system. The AC dynamics of this system will be dominated by the complex dynamics of Voltage Source Converters (VSCs). \Cref{fig:VSC_Control} shows the inner control loops on the AC side of a typical VSC~\cite{Misyris:2020_NSWPH,Bastin:2019}. The active and reactive power controllers associated with this VSC are depicted in greater detail in \cref{fig:NSWPH_Control}. In particular, these power controllers explicitly depend on droop parameters which determine the system's participation in both primary frequency and primary voltage control ($K_{p,f}$ and $K_v$, respectively). The influence of these parameters inside of the active and reactive power VSC control loops is depicted in \cref{fig:NSWPH_Control}. 

As shown, the control loops also depend on active and reactive power reference set points as input parameters. In the optimal dispatch problem, we assume grid operators have simultaneous control over $P_{\rm ref}$, $Q_{\rm ref}$, $K_{p,f}$ and $K_v$. Notably, $K_{p,f}$ and $K_v$ only affect the \textit{dynamics} of the NSWPH system (rather than its steady state behavior). The active and reactive power set points, however, affect the power flow initialization of the entire circuit, thus implicitly altering the system's linearized dynamics.

When dispatching the power set points and the quantity of control regulation provided by this system, operators must ensure that local system damping ratios remain above some minimally acceptable threshold (e.g., 3\%). The damping ratio $\zeta_i$ associated with the $i^{\rm th}$ complex eigenvalue $\lambda_i=\sigma_i + j\omega_i$ is given by 
\begin{align}
\zeta_i=\frac{-100\times\sigma_i}{\sqrt{\sigma_i^2 + \omega_i^2}}.
\end{align}
In this paper, we assume this margin has to be maintained\footnote{We enforce this damping ratio constraint across all low frequency ($<500$ Hz) eigenmodes. Several high frequency (kHz-range) eigenmodes in this system remain closer to $j\omega$ axis.} across all credible N-1 contingencies (i.e., the loss of any one turbine). The mapping between the input  parameters $P_{\rm ref}$, $Q_{\rm ref}$, $K_{p,f}$ and $K_v$ and the minimum system damping ratio, however, is intractably dense; for any change in their values, the mapping involves (i) re-solving the network AC power flow problem, (ii) initializing all state variables, (iii) linearizing around the new operating point, (iv) eigendecomposition of the associated state matrix, (v) explicit computation of the minimum damping ratios, and (iv) repetition of this process across all N-1 contingencies.

\subsection{Directed Walk Procedure}\label{App_DW}
To perform the directed walk, we follow a procedure similar to the one outlined in~\cite{Thams2020}. For each operating point, we compute the sensitivity of the $n^{\rm th}$ eigenvalue to the $i^{\rm th}$ element $\bm{x}_{i}$ of the operating point~\cite{Thams2020}:
\begin{align}\label{eq: eig_sensitive}
\frac{\partial\lambda_{n}}{\partial\bm{x}_{i}} & \triangleq\frac{\partial\sigma_{n}}{\partial\bm{x}_{i}}+j\frac{\partial\omega_{n}}{\partial\bm{x}_{i}}=\psi_{n}^{T}\frac{A_{\bm{x}}}{\partial\bm{x}_{i}}\phi_{n},
\end{align}
where $\psi_{n}^{T}$ and $\phi$ are the left and right eigenvectors, normalized such that $\psi_{n}^{T}\phi_{n}=1,\forall n$, and $A_{\bm{x}}$ is the state matrix of the NSWPH in Fig. \ref{fig:NSWPH}, linearized around operating point $\bm x$. Using the real and imaginary sensitivities in (\ref{eq: eig_sensitive}), damping ratio sensitivity is given as
\begin{align}\label{eq: sensitivity}
\frac{\partial\zeta_{n}}{\partial\bm{x}_{i}}=\omega_{n}\frac{\sigma_{n}\frac{\partial\omega_{n}}{\partial\bm{x}_{i}}-\omega_{n}\frac{\partial\sigma_{n}}{\partial\bm{x}_{i}}}{(\sigma_{n}^{2}+\omega_{n}^{2})^{\frac{3}{2}}}\triangleq\nabla_{\bm{x}_{i}}.
\end{align}
The concatenated gradient information $\nabla_{\bm{x}}$ allows us to push the operating point $\bm{x}$ towards the stability boundary via steepest descent:
\begin{align}\label{eq: SD}
\bm{x}^{(i+1)}=\bm{x}^{(i)}+\alpha(\bm{x}^{(i)})\nabla_{\bm{x}^{(i)}},
\end{align}
where the magnitude of the stepsize $\alpha(\bm{x}^{(i)})$ is adaptively chosen based on the distance to the stability margin, and its sign ($+/-$) is chosen based on which side of the stability margin the operating point $\bm{x}^{(i)}$ is on.

In this work, we utilize a descent procedure which holds control parameters ($K_{p,f}$, $K_v$) constant while the power set points ($P_{\rm ref}$, $Q_{\rm ref}$) are iteratively updated. Once the stability margin has been reached, the initial operating point is returned to, and the procedure is reversed (power is held constant while control parameters walk).

\subsection{Detailed numerical results}

\begin{table}[ht]
  \caption{Selected hyper-parameter value}
  \label{tbl:hyperparameter_values}
  \centering
  \begin{tabular}{lcccccc}
    \toprule
    Model/dataset & $K$ & $N_k$ & $l_0$ & $\gamma$ & $\alpha_J$ & $\Delta$ \\
    \midrule
    Grid 5 & 3& 32 & 0.05 & 1 & - & - \\
    Grid 6 & 4 & 64 & 0.02 & 1 & - & - \\
    Grid 7 & 3 & 64 & 0.02 & 1 & - & - \\
    Grid 5 + DW & 4 & 32 & 0.05 & 1 & - & - \\
    Grid 5 + NI & 4 & 32 & 0.05 & 1 & - & 3.0\% \\
    Grid 5 + VI & 3 & 16 & 0.05 & 1 & - & 0.25\% \\
    Grid 5 + PR & 4 & 64 & 0.02 & 1 & 0.1 & - \\
    \midrule
    LHC 5 & 4 & 32 & 0.05 & 1 & - & - \\
    LHC 6 & 3 & 32 & 0.05 & 1 & - & -\\
    LHC 7 & 3 & 64 & 0.02 & 1 & - & -\\
    LHC 5 + DW & 4 & 32 & 0.05 & 1 & - & -\\
    LHC 5 + NI & 2 & 32 & 0.05 & 1 & - & 3.0\%\\
    LHC 5 + VI & 3 & 32 & 0.05 & 1 & - & 0.25\% \\
    LHC 5 + PR & 4 & 64 & 0.01 & 1 & 0.1 & - \\
    \midrule
    Uniform 5 & 4 & 32 & 0.05 & 1 & - & \\
    Uniform 6 & 4 & 32 & 0.05 & 1 & - & \\
    Uniform 7 & 4 & 32 & 0.05 & 1 & & \\
    Uniform 5 + DW & 3 & 32 & 0.05 & 1 & - & - \\
    Uniform 5 + NI & 3 & 64 & 0.02 & 1 & - & 0.25\% \\
    Uniform 5 + VI & 3 & 32 & 0.05 & 1 & - & 3.0\% \\
    Uniform 5 + PR & 3 & 32 & 0.02 & 1 & 0.1 & - \\
    \bottomrule
  \end{tabular}

\end{table}

\begin{table*}[ht]
  \caption{
  \scriptsize Sizes of the stability classes in the used datasets to illustrate the class imbalance. For the versions with resampling (+ DW, + NI, and + VI) only the additionally sampled points are reported. Due to the dependency on the initial training, the added points for the dataset + NI, + VI vary for different training runs, hence the mean and standard deviation of which share belongs to each class are shown.}
  \label{tbl:dataset_classes}
  \centering
  \begin{tabular}{lcccccc}
    \toprule
    Dataset & N & Stable & marg. stable & marginal & marg. unstable & unstable \\
    \midrule
    Test & 194481 & 81.4\% & 4.67\% & 0.72\% & 3.35\% & 9.83\%\\
    \midrule
    Grid 5 & 625 & 74.88\% & 5.12\% & 1.12\% & 3.68\% & 15.20\% \\
    Grid 6 & 1296 & 75.85\% & 6.12\% & 0.46\% & 3.63\% & 13.97\% \\
    Grid 7 & 2401 & 77.09\% & 5.58\% & 0.96\% & 3.67\% & 12.70\% \\
    \midrule
    Grid 5 + DW & 202 & 3.47\% & 16.86\% & 76.73\% & 2.48\% & 0.50\% \\
    Grid 5 + NI & 200 & 9.34 $\pm$ 3.64\% & 39.97 $\pm$ 4.16\% & 8.62 $\pm$ 1.90\% & 37.92 $\pm$ 3.78\% & 4.16 $\pm$ 2.17\%\\
    Grid 5 + VI & 200 & 73.67 $\pm$ 3.07\% & 7.49 $\pm$ 2.12\% & 1.18 $\pm$ 0.83\% & 5.37 $\pm$ 1.48\% & 12.30 $\pm$ 2.22\% \\
    \midrule
    LHC 5 & 625 & 82.40\% & 4.96\% & 0.48\% & 2.88\% & 9.28\% \\
    LHC 6 & 1296 & 83.33\% & 4.32\% & 0.54\% & 3.32\% & 8.49\% \\
    LHC 7 & 2401 & 83.17\% & 4.50\% & 0.67\% & 3.54\% & 8.12\% \\ 
    \midrule
    LHC 5 + DW & 156 & 3.21 \% & 14.10\% & 82.05\% & 0.64\% & 0.00\% \\
    LHC 5 + NI & 200 & 5.91 $\pm$ 2.37\% & 43.50 $\pm$ 3.94\% & 8.68 $\pm$ 1.96\% & 37.48 $\pm$ 3.71\% & 4.44 $\pm$ 1.73\%\\
    LHC 5 + VI & 200 & 68.39 $\pm$ 3.03\% & 8.64 $\pm$ 2.17\% & 1.47 $\pm$ 0.82\% & 6.04 $\pm$ 1.62\% & 15.46 $\pm$ 2.82\%\\
    \midrule
    Uniform 5 & 625 & 82.24\% & 4.96\% & 0.96\% & 2.56\% & 9.28\% \\
    Uniform 6 & 1296 & 83.26\% & 5.40\% & 0.77\% & 2.70\% & 7.87\% \\
    Uniform 7 & 2401 & 83.92\% & 4.29\% & 0.79\% & 2.75\% & 8.25\% \\
    \midrule
    Uniform 5 + DW & 182 & 3.85 \% & 13.74\% & 82.42\% & 0.00\% & 0.00\%\\
    Uniform 5 + NI & 200 & 0.65 $\pm$ 0.83\% & 29.80 $\pm$ 7.08\% & 24.07 $\pm$ 4.99\% & 44.59 $\pm$ 9.03\% & 0.89 $\pm$ 1.10\%\\
    Uniform 5 + VI & 200 & 73.91 $\pm$ 3.23\% & 7.27 $\pm$ 1.85\% & 1.06 $\pm$ 0.74\% & 5.13 $\pm$ 1.62\% & 12.62 $\pm$ 2.45\%\\
    \bottomrule
  \end{tabular}
\end{table*}

\begin{table*}[ht]
  \caption{\scriptsize Performance dependent on the sampling technique (grid, latin hypercube sampling (LHC), or uniform) with different dataset size ($5^4$, $6^4$, $7^4$), with directed walks (DW), with neural network informed points (NI), with verification informed points (VI), and with physics-regularization (PR).}
  \label{tbl:error_comparison}
  \centering
  \begin{tabular}{llcccccc}
    \toprule
     & & & \multicolumn{5}{c}{Mean squared error per classification ($\mathcal{L}_y$ in \eqref{eq:data_loss})}\\
    \cmidrule(lr){4-8}
    Model/dataset & Dataset size $N$ &  MSE total & Stable & marg. stable & marginal & marg. unstable & unstable\\
    \midrule
    Grid 5 & 625 & 0.933 $\pm$ 0.325 & 0.790 $\pm$ 0.234 & 2.794 $\pm$ 1.324 & 2.633 $\pm$ 1.640 & 2.265 $\pm$ 1.753 & 0.655 $\pm$ 0.363\\
    Grid 6 & 1296 & 0.487 $\pm$ 0.070 & 0.481 $\pm$ 0.077 & 0.957 $\pm$ 0.210 & 0.831 $\pm$ 0.185 & 0.723 $\pm$ 0.164 & 0.211 $\pm$ 0.051\\
    Grid 7 & 2401 & 0.277 $\pm$ 0.049 & 0.288 $\pm$ 0.054 & 0.416 $\pm$ 0.122 & 0.356 $\pm$ 0.103 & 0.316 $\pm$ 0.089 & 0.108 $\pm$ 0.027\\
    \midrule
    Grid 5 + DW & 827 & 0.741 $\pm$ 0.174 & 0.781 $\pm$ 0.195 & 0.968 $\pm$ 0.361 & 0.679 $\pm$ 0.262 & 0.630 $\pm$ 0.240 & 0.341 $\pm$ 0.113\\
    Grid 5 + NI & 825 & 0.666 $\pm$ 0.210 & 0.620 $\pm$ 0.170 & 1.712 $\pm$ 1.155 & 1.281 $\pm$ 1.031 & 0.985 $\pm$ 0.831 & 0.395 $\pm$ 0.190\\
    Grid 5 + VI & 825 & 0.685 $\pm$ 0.215 & 0.623 $\pm$ 0.194 & 1.605 $\pm$ 0.838 & 1.466 $\pm$ 0.795 & 1.269 $\pm$ 0.673 & 0.508 $\pm$ 0.222\\
    Grid 5 + PR & 625 & 0.680 $\pm$ 0.101 & 0.611 $\pm$ 0.095 & 1.750 $\pm$ 0.520 & 1.727 $\pm$ 0.556 & 1.530 $\pm$ 0.495 & 0.375 $\pm$ 0.107\\
    \midrule
    \midrule
    LHC 5 & 625 & 0.698 $\pm$ 0.174 & 0.470 $\pm$ 0.125 & 2.212 $\pm$ 0.565 & 2.265 $\pm$ 0.624 & 2.502 $\pm$ 0.617 & 1.141 $\pm$ 0.375\\
    LHC 6 & 1296 & 0.391 $\pm$ 0.286 & 0.245 $\pm$ 0.154 & 1.240 $\pm$ 1.245 & 1.285 $\pm$ 1.722 & 1.548 $\pm$ 1.875 & 0.737 $\pm$ 0.314\\
    LHC 7 & 2401 & 0.261 $\pm$ 0.025 & 0.135 $\pm$ 0.017 & 0.936 $\pm$ 0.108 & 0.995 $\pm$ 0.121 & 1.286 $\pm$ 0.127 & 0.578 $\pm$ 0.060\\
    \midrule
    LHC 5 + DW & 781 & 0.546 $\pm$ 0.134 & 0.391 $\pm$ 0.140 & 1.427 $\pm$ 0.243 & 1.356 $\pm$ 0.201 & 1.732 $\pm$ 0.193 & 0.948 $\pm$ 0.166\\
    LHC 5 + NI & 825 & 0.502 $\pm$ 0.090 & 0.369 $\pm$ 0.084 & 1.270 $\pm$ 0.233 & 1.326 $\pm$ 0.224 & 1.600 $\pm$ 0.241 & 0.804 $\pm$ 0.142\\
    LHC 5 + VI & 825 & 0.615 $\pm$ 0.472 & 0.385 $\pm$ 0.233 & 2.132 $\pm$ 2.247 & 2.339 $\pm$ 3.088 & 2.645 $\pm$ 3.343 & 0.981 $\pm$ 0.520\\
    LHC 5 + PR & 625 & 0.378 $\pm$ 0.037 & 0.258 $\pm$ 0.035 & 1.116 $\pm$ 0.139 & 1.180 $\pm$ 0.138 & 1.446 $\pm$ 0.142 & 0.602 $\pm$ 0.065\\
    \midrule
    \midrule
    Uniform 5 & 625 & 0.757 $\pm$ 0.316 & 0.578 $\pm$ 0.196 & 1.889 $\pm$ 1.237 & 1.936 $\pm$ 1.688 & 2.155 $\pm$ 1.868 & 1.140 $\pm$ 0.539\\
    Uniform 6 & 1296 & 0.443 $\pm$ 0.090 & 0.264 $\pm$ 0.068 & 1.489 $\pm$ 0.418 & 1.631 $\pm$ 0.454 & 1.892 $\pm$ 0.381 & 0.846 $\pm$ 0.175\\
    Uniform 7 & 2401 & 0.288 $\pm$ 0.042 & 0.159 $\pm$ 0.037 & 1.022 $\pm$ 0.120 & 1.091 $\pm$ 0.111 & 1.310 $\pm$ 0.110 & 0.601 $\pm$ 0.086\\
    \midrule
    Uniform 5 + DW & 807 & 0.657 $\pm$ 0.128 & 0.601 $\pm$ 0.125 & 0.933 $\pm$ 0.238 & 0.860 $\pm$ 0.249 & 1.080 $\pm$ 0.324 & 0.829 $\pm$ 0.319\\
    Uniform 5 + NI & 825 & 0.549 $\pm$ 0.093 & 0.440 $\pm$ 0.081 & 1.134 $\pm$ 0.312 & 1.191 $\pm$ 0.336 & 1.453 $\pm$ 0.355 & 0.819 $\pm$ 0.202\\
    Uniform 5 + VI & 825& 0.559 $\pm$ 0.287 & 0.407 $\pm$ 0.161 & 1.477 $\pm$ 1.277 & 1.564 $\pm$ 1.746 & 1.788 $\pm$ 1.903 & 0.889 $\pm$ 0.366\\
    Uniform 5 + PR & 625 & 0.332 $\pm$ 0.040 & 0.298 $\pm$ 0.041 & 0.545 $\pm$ 0.138 & 0.573 $\pm$ 0.177 & 0.663 $\pm$ 0.222 & 0.382 $\pm$ 0.093\\
    \bottomrule
  \end{tabular}
\end{table*}

\label{sec:appendix}







\bibliographystyle{IEEEtran}
\bibliography{references.bib}

\begin{thebibliography}{10}
\providecommand{\url}[1]{#1}
\csname url@samestyle\endcsname
\providecommand{\newblock}{\relax}
\providecommand{\bibinfo}[2]{#2}
\providecommand{\BIBentrySTDinterwordspacing}{\spaceskip=0pt\relax}
\providecommand{\BIBentryALTinterwordstretchfactor}{4}
\providecommand{\BIBentryALTinterwordspacing}{\spaceskip=\fontdimen2\font plus
\BIBentryALTinterwordstretchfactor\fontdimen3\font minus
  \fontdimen4\font\relax}
\providecommand{\BIBforeignlanguage}[2]{{%
\expandafter\ifx\csname l@#1\endcsname\relax
\typeout{** WARNING: IEEEtran.bst: No hyphenation pattern has been}%
\typeout{** loaded for the language `#1'. Using the pattern for}%
\typeout{** the default language instead.}%
\else
\language=\csname l@#1\endcsname
\fi
#2}}
\providecommand{\BIBdecl}{\relax}
\BIBdecl

\bibitem{Misyris:2020_NSWPH}
G.~{Misyris}, T.~{Van Cutsem}, J.~{M{\o}ller}, M.~{Dijokas}, O.~{Renom
  Estragu{\'e}s}, B.~{Bastin}, S.~{Chatzivasileiadis}, A.~{Nielsen},
  T.~{Weckesser}, J.~{{\O}stergaard}, and F.~{Kryezi}, ``{North Sea Wind Power
  Hub: System Configurations, Grid Implementation and Techno-economic
  Assessment},'' \emph{arXiv e-prints}, p. arXiv:2006.05829, Jun. 2020.

\bibitem{Amin:2021}
M.~R. Amin, M.~Negnevitsky, E.~Franklin, K.~S. Alam, and S.~B. Naderi,
  ``Application of battery energy storage systems for primary frequency control
  in power systems with high renewable energy penetration,'' \emph{Energies},
  vol.~14, no.~5, p. 1379, Mar 2021.

\bibitem{KhareSaxena:2020}
A.~KhareSaxena, S.~Saxena, and K.~Sudhakar, ``Solar energy policy of india: An
  overview,'' \emph{CSEE Journal of Power and Energy Systems}, pp. 1--32, 2020.

\bibitem{Kody:2021}
A.~{Kody}, S.~{Chevalier}, S.~{Chatzivasileiadis}, and D.~{Molzahn},
  ``{Modeling the AC Power Flow Equations with Optimally Compact Neural
  Networks: Application to Unit Commitment},'' \emph{arXiv e-prints}, p.
  arXiv:2110.11269, Oct. 2021.

\bibitem{Murzakhanov:2020}
I.~{Murzakhanov}, A.~{Venzke}, G.~S. {Misyris}, and S.~{Chatzivasileiadis},
  ``{Neural Networks for Encoding Dynamic Security-Constrained Optimal Power
  Flow},'' \emph{arXiv e-prints}, p. arXiv:2003.07939, Mar. 2020.

\bibitem{Duchesne2020}
L.~Duchesne, E.~Karangelos, and L.~Wehenkel, ``Recent developments in machine
  learning for energy systems reliability management,'' \emph{Proceedings of
  the IEEE}, vol. 108, no.~9, pp. 1656--1676, 2020.

\bibitem{Marot:2020}
A.~Marot, B.~Donnot, C.~Romero, B.~Donon, M.~Lerousseau, L.~Veyrin-Forrer, and
  I.~Guyon, ``Learning to run a power network challenge for training topology
  controllers,'' \emph{Electric Power Systems Research}, vol. 189, p. 106635,
  2020.

\bibitem{Donnot:2017}
B.~{Donnot}, I.~{Guyon}, M.~{Schoenauer}, P.~{Panciatici}, and A.~{Marot},
  ``{Introducing machine learning for power system operation support},''
  \emph{arXiv e-prints}, p. arXiv:1709.09527, Sep. 2017.

\bibitem{Stiasny_github:2022}
J.~Stiasny, ``Closing the loop: A framework for trustworthy machine learning in
  power systems,'' \url{https://github.com/jbesty/irep_2022_closing_the_loop},
  2022.

\bibitem{Venzke:2021}
A.~Venzke, D.~K. Molzahn, and S.~Chatzivasileiadis, ``Efficient creation of
  datasets for data-driven power system applications,'' \emph{Electric Power
  Systems Research}, vol. 190, p. 106614, 2021.

\bibitem{Thams2020}
F.~Thams, A.~Venzke, R.~Eriksson, and S.~Chatzivasileiadis, ``Efficient
  database generation for data-driven security assessment of power systems,''
  \emph{IEEE Transactions on Power Systems}, vol.~35, no.~1, pp. 30--41, 2020.

\bibitem{Genc:2010}
I.~Genc, R.~Diao, V.~Vittal, S.~Kolluri, and S.~Mandal, ``Decision tree-based
  preventive and corrective control applications for dynamic security
  enhancement in power systems,'' \emph{IEEE Transactions on Power Systems},
  vol.~25, no.~3, pp. 1611--1619, 2010.

\bibitem{Vincent:2017}
V.~Tjeng, K.~Xiao, and R.~Tedrake, ``Evaluating robustness of neural networks
  with mixed integer programming,'' \emph{arXiv preprint arXiv:1711.07356},
  2017.

\bibitem{Goodfellow:2016}
I.~Goodfellow, Y.~Bengio, and A.~Courville, \emph{Deep learning}.\hskip 1em
  plus 0.5em minus 0.4em\relax MIT press, 2016.

\bibitem{Hamon:2016}
C.~Hamon, M.~Perninge, and L.~Söder, ``An importance sampling technique for
  probabilistic security assessment in power systems with large amounts of wind
  power,'' \emph{Electric Power Systems Research}, vol. 131, pp. 11--18, 2016.

\bibitem{Krishnan:2011}
V.~Krishnan, J.~D. McCalley, S.~Henry, and S.~Issad, ``Efficient database
  generation for decision tree based power system security assessment,''
  \emph{IEEE Transactions on Power Systems}, vol.~26, no.~4, pp. 2319--2327,
  2011.

\bibitem{Molzahn:2017}
D.~K. Molzahn, ``Computing the feasible spaces of optimal power flow
  problems,'' \emph{IEEE Transactions on Power Systems}, vol.~32, no.~6, pp.
  4752--4763, 2017.

\bibitem{Coffrin:2016}
C.~Coffrin, H.~L. Hijazi, and P.~Van~Hentenryck, ``The qc relaxation: A
  theoretical and computational study on optimal power flow,'' \emph{IEEE
  Transactions on Power Systems}, vol.~31, no.~4, pp. 3008--3018, 2016.

\bibitem{Kaufman:1998}
D.~E. Kaufman and R.~L. Smith, ``Direction choice for accelerated convergence
  in hit-and-run sampling,'' \emph{Operations Research}, vol.~46, no.~1, pp.
  84--95, 1998.

\bibitem{Jones:2021}
T.~{Joswig-Jones}, K.~{Baker}, and A.~S. {Zamzam}, ``{OPF-Learn: An Open-Source
  Framework for Creating Representative AC Optimal Power Flow Datasets},''
  \emph{arXiv e-prints}, p. arXiv:2111.01228, Nov. 2021.

\bibitem{Wehenkel:1994}
L.~Wehenkel, M.~Pavella, E.~Euxibie, and B.~Heilbronn, ``Decision tree based
  transient stability method a case study,'' \emph{IEEE Transactions on Power
  Systems}, vol.~9, no.~1, pp. 459--469, 1994.

\bibitem{Hatziargyriou:1994}
N.~Hatziargyriou, G.~Contaxis, and N.~Sideris, ``A decision tree method for
  on-line steady state security assessment,'' \emph{IEEE Transactions on Power
  Systems}, vol.~9, no.~2, pp. 1052--1061, 1994.

\bibitem{cremer_machine-learning_2021}
J.~L. Cremer and G.~Strbac, ``A machine-learning based probabilistic
  perspective on dynamic security assessment,'' \emph{International Journal of
  Electrical Power \& Energy Systems}, vol. 128, p. 106571, 2021.

\bibitem{Venzke:2020}
A.~Venzke, G.~Qu, S.~Low, and S.~Chatzivasileiadis, ``Learning optimal power
  flow: Worst-case guarantees for neural networks,'' in \emph{2020 IEEE
  International Conference on Communications, Control, and Computing
  Technologies for Smart Grids (SmartGridComm)}.\hskip 1em plus 0.5em minus
  0.4em\relax IEEE, 2020, pp. 1--7.

\bibitem{montavon_neural_2012}
G.~Montavon, G.~B. Orr, and K.-R. Müller, Eds.,
  \emph{\BIBforeignlanguage{en}{Neural {Networks}: {Tricks} of the {Trade}}},
  2nd~ed., ser. Lecture {Notes} in {Computer} {Science}.\hskip 1em plus 0.5em
  minus 0.4em\relax Berlin, Heidelberg: Springer Berlin Heidelberg, 2012, vol.
  7700.

\bibitem{hastie_elements_2009}
T.~Hastie, R.~Tibshirani, and J.~Friedman, \emph{The {Elements} of
  {Statistical} {Learning}}, ser. Springer {Series} in {Statistics}.\hskip 1em
  plus 0.5em minus 0.4em\relax New York, NY: Springer New York, 2009.

\bibitem{blum_selection_1997}
A.~L. Blum and P.~Langley, ``Selection of relevant features and examples in
  machine learning,'' \emph{Artificial Intelligence}, vol.~97, no.~1, pp.
  245--271, 1997.

\bibitem{guyon_introduction_2003}
I.~Guyon and A.~Elisseeff, ``An introduction to variable and feature
  selection,'' \emph{J. Mach. Learn. Res.}, vol.~3, no. null, p. 1157–1182,
  2003.

\bibitem{salcedo-sanz_feature_2018}
S.~Salcedo-Sanz, L.~Cornejo-Bueno, L.~Prieto, D.~Paredes, and
  R.~García-Herrera, ``Feature selection in machine learning prediction
  systems for renewable energy applications,'' \emph{Renewable and Sustainable
  Energy Reviews}, vol.~90, pp. 728--741, 2018.

\bibitem{Kahloot:2021}
K.~M. Kahloot and P.~Ekler, ``Algorithmic splitting: A method for dataset
  preparation,'' \emph{IEEE Access}, vol.~9, pp. 125\,229--125\,237, 2021.

\bibitem{Brunton:2019}
S.~L. Brunton and J.~N. Kutz, \emph{Data-driven science and engineering:
  Machine learning, dynamical systems, and control}.\hskip 1em plus 0.5em minus
  0.4em\relax Cambridge University Press, 2019.

\bibitem{baydin_automatic_2018}
A.~G. Baydin, B.~A. Pearlmutter, A.~A. Radul, and J.~M. Siskind, ``Automatic
  differentiation in machine learning: a survey,'' \emph{Journal of Machine
  Learning Research}, vol.~18, no. 153, pp. 1--43, 2018.

\bibitem{Raissi:2019}
M.~Raissi, P.~Perdikaris, and G.~Karniadakis, ``Physics-informed neural
  networks: A deep learning framework for solving forward and inverse problems
  involving nonlinear partial differential equations,'' \emph{Journal of
  Computational Physics}, vol. 378, pp. 686--707, 2019.

\bibitem{Cuomo:2022}
S.~Cuomo, V.~S. Di~Cola, F.~Giampaolo, G.~Rozza, M.~Raissi, and F.~Piccialli,
  ``Scientific machine learning through physics-informed neural networks: Where
  we are and what's next,'' \emph{arXiv preprint arXiv:2201.05624}, 2022.

\bibitem{Misyris:2020}
G.~S. Misyris, A.~Venzke, and S.~Chatzivasileiadis, ``Physics-{Informed}
  {Neural} {Networks} for {Power} {Systems},'' in \emph{2020 {IEEE} {Power} \&
  {Energy} {Society} {General} {Meeting}}, Aug. 2020, pp. 1--5.

\bibitem{Stiasny:2021}
J.~Stiasny, G.~S. Misyris, and S.~Chatzivasileiadis, ``Physics-informed neural
  networks for non-linear system identification for power system dynamics,'' in
  \emph{2021 IEEE Powertech}, 2021, pp. 1--7.

\bibitem{stiasny_transient_2021}
------, ``Transient {Stability} {Analysis} with {Physics}-{Informed} {Neural}
  {Networks},'' \emph{arXiv:2106.13638}, Jun. 2021.

\bibitem{Stiasny:2021_Learning}
J.~Stiasny, S.~Chevalier, and S.~Chatzivasileiadis, ``Learning without data:
  Physics-informed neural networks for fast time-domain simulation,'' in
  \emph{2021 IEEE International Conference on Communications, Control, and
  Computing Technologies for Smart Grids (SmartGridComm)}.\hskip 1em plus 0.5em
  minus 0.4em\relax IEEE, 2021, pp. 438--443.

\bibitem{Singh:2021}
M.~Singh, V.~Kekatos, and G.~B. Giannakis, ``Learning to solve the ac-opf using
  sensitivity-informed deep neural networks,'' \emph{IEEE Transactions on Power
  Systems}, 2021.

\bibitem{Singh:2020}
M.~K. Singh, S.~Gupta, V.~Kekatos, G.~Cavraro, and A.~Bernstein, ``Learning to
  optimize power distribution grids using sensitivity-informed deep neural
  networks,'' in \emph{2020 IEEE International Conference on Communications,
  Control, and Computing Technologies for Smart Grids (SmartGridComm)}, 2020,
  pp. 1--6.

\bibitem{Nellikkath2:2021}
R.~Nellikkath and S.~Chatzivasileiadis, ``Physics-informed neural networks for
  ac optimal power flow,'' \emph{arXiv preprint arXiv:2110.02672}, 2021.

\bibitem{Nellikkath:2021}
------, ``Physics-informed neural networks for minimising worst-case violations
  in dc optimal power flow,'' in \emph{2021 IEEE International Conference on
  Communications, Control, and Computing Technologies for Smart Grids
  (SmartGridComm)}.\hskip 1em plus 0.5em minus 0.4em\relax IEEE, 2021, pp.
  419--424.

\bibitem{Pagnier:2021}
L.~Pagnier and M.~Chertkov, ``Physics-{Informed} {Graphical} {Neural} {Network}
  for {Parameter} \& {State} {Estimations} in {Power} {Systems},''
  \emph{arXiv:2102.06349}, Feb. 2021.

\bibitem{Lin:2021}
S.~{Lin}, S.~{Liu}, and H.~{Zhu}, ``{Risk-Aware Learning for Scalable Voltage
  Optimization in Distribution Grids},'' \emph{arXiv e-prints}, p.
  arXiv:2110.01490, Oct. 2021.

\bibitem{Xiang:2018}
W.~{Xiang}, P.~{Musau}, A.~A. {Wild}, D.~{Manzanas Lopez}, N.~{Hamilton},
  X.~{Yang}, J.~{Rosenfeld}, and T.~T. {Johnson}, ``{Verification for Machine
  Learning, Autonomy, and Neural Networks Survey},'' \emph{arXiv e-prints}, p.
  arXiv:1810.01989, Oct. 2018.

\bibitem{vovk_algorithmic_2005}
V.~Vovk, A.~Gammerman, and G.~Shafer, \emph{Algorithmic learning in a random
  world}.\hskip 1em plus 0.5em minus 0.4em\relax Springer, 2005.

\bibitem{tibshirani_conformal_2019}
R.~J. Tibshirani, R.~Foygel~Barber, E.~Candes, and A.~Ramdas, ``Conformal
  prediction under covariate shift,'' in \emph{Advances in Neural Information
  Processing Systems}, vol.~32, 2019.

\bibitem{Wong:2018}
E.~Wong and Z.~Kolter, ``Provable defenses against adversarial examples via the
  convex outer adversarial polytope,'' in \emph{International Conference on
  Machine Learning}.\hskip 1em plus 0.5em minus 0.4em\relax PMLR, 2018, pp.
  5286--5295.

\bibitem{Dvijotham:2018}
K.~Dvijotham, R.~Stanforth, S.~Gowal, T.~A. Mann, and P.~Kohli, ``A dual
  approach to scalable verification of deep networks.'' in \emph{UAI}, vol.~1,
  no.~2, 2018, p.~3.

\bibitem{Dathathri:2020}
S.~Dathathri, K.~Dvijotham, A.~Kurakin, A.~Raghunathan, J.~Uesato, R.~R. Bunel,
  S.~Shankar, J.~Steinhardt, I.~Goodfellow, P.~S. Liang \emph{et~al.},
  ``Enabling certification of verification-agnostic networks via
  memory-efficient semidefinite programming,'' \emph{Advances in Neural
  Information Processing Systems}, vol.~33, pp. 5318--5331, 2020.

\bibitem{Fazlyab:2020}
M.~Fazlyab, M.~Morari, and G.~J. Pappas, ``Safety verification and robustness
  analysis of neural networks via quadratic constraints and semidefinite
  programming,'' \emph{IEEE Transactions on Automatic Control}, 2020.

\bibitem{Dvijotham:2020}
K.~D. Dvijotham, R.~Stanforth, S.~Gowal, C.~Qin, S.~De, and P.~Kohli,
  ``Efficient neural network verification with exactness characterization,'' in
  \emph{Uncertainty in artificial intelligence}.\hskip 1em plus 0.5em minus
  0.4em\relax PMLR, 2020, pp. 497--507.

\bibitem{Mirman:2018}
M.~Mirman, T.~Gehr, and M.~Vechev, ``Differentiable abstract interpretation for
  provably robust neural networks,'' in \emph{International Conference on
  Machine Learning}.\hskip 1em plus 0.5em minus 0.4em\relax PMLR, 2018, pp.
  3578--3586.

\bibitem{Gehr:2018}
T.~Gehr, M.~Mirman, D.~Drachsler-Cohen, P.~Tsankov, S.~Chaudhuri, and
  M.~Vechev, ``Ai2: Safety and robustness certification of neural networks with
  abstract interpretation,'' in \emph{2018 IEEE Symposium on Security and
  Privacy (SP)}.\hskip 1em plus 0.5em minus 0.4em\relax IEEE, 2018, pp. 3--18.

\bibitem{Weng:2018}
L.~Weng, H.~Zhang, H.~Chen, Z.~Song, C.-J. Hsieh, L.~Daniel, D.~Boning, and
  I.~Dhillon, ``Towards fast computation of certified robustness for relu
  networks,'' in \emph{International Conference on Machine Learning}.\hskip 1em
  plus 0.5em minus 0.4em\relax PMLR, 2018, pp. 5276--5285.

\bibitem{Zhang:2018}
H.~Zhang, T.-W. Weng, P.-Y. Chen, C.-J. Hsieh, and L.~Daniel, ``Efficient
  neural network robustness certification with general activation functions,''
  \emph{Advances in neural information processing systems}, vol.~31, 2018.

\bibitem{Hein:2017}
M.~Hein and M.~Andriushchenko, ``Formal guarantees on the robustness of a
  classifier against adversarial manipulation,'' \emph{Advances in neural
  information processing systems}, vol.~30, 2017.

\bibitem{Wang3:2018}
S.~Wang, K.~Pei, J.~Whitehouse, J.~Yang, and S.~Jana, ``Formal security
  analysis of neural networks using symbolic intervals,'' in \emph{27th USENIX
  Security Symposium (USENIX Security 18)}, 2018, pp. 1599--1614.

\bibitem{Garcia:2015}
J.~Garc{\i}a and F.~Fern{\'a}ndez, ``A comprehensive survey on safe
  reinforcement learning,'' \emph{Journal of Machine Learning Research},
  vol.~16, no.~1, pp. 1437--1480, 2015.

\bibitem{Berkenkamp:2017}
F.~Berkenkamp, M.~Turchetta, A.~Schoellig, and A.~Krause, ``Safe model-based
  reinforcement learning with stability guarantees,'' \emph{Advances in neural
  information processing systems}, vol.~30, 2017.

\bibitem{Gillula:2012}
J.~H. Gillula and C.~J. Tomlin, ``Guaranteed safe online learning via
  reachability: tracking a ground target using a quadrotor,'' in \emph{2012
  IEEE International Conference on Robotics and Automation}.\hskip 1em plus
  0.5em minus 0.4em\relax IEEE, 2012, pp. 2723--2730.

\bibitem{Dutta:2018}
S.~Dutta, S.~Jha, S.~Sankaranarayanan, and A.~Tiwari, ``Output range analysis
  for deep feedforward neural networks,'' in \emph{NASA Formal Methods
  Symposium}.\hskip 1em plus 0.5em minus 0.4em\relax Springer, 2018, pp.
  121--138.

\bibitem{Lomuscio:2017}
A.~Lomuscio and L.~Maganti, ``An approach to reachability analysis for
  feed-forward relu neural networks,'' \emph{arXiv preprint arXiv:1706.07351},
  2017.

\bibitem{Xiao:2018}
K.~Y. Xiao, V.~Tjeng, N.~M. Shafiullah, and A.~Madry, ``Training for faster
  adversarial robustness verification via inducing relu stability,''
  \emph{arXiv preprint arXiv:1809.03008}, 2018.

\bibitem{Venzke2:2020}
A.~Venzke and S.~Chatzivasileiadis, ``Verification of neural network behaviour:
  Formal guarantees for power system applications,'' \emph{IEEE Transactions on
  Smart Grid}, vol.~12, no.~1, pp. 383--397, 2020.

\bibitem{Grimstad:2019}
B.~{Grimstad} and H.~{Andersson}, ``{ReLU Networks as Surrogate Models in
  Mixed-Integer Linear Programs},'' \emph{arXiv e-prints}, p. arXiv:1907.03140,
  Jul. 2019.

\bibitem{Martinez:2017}
N.~Martinez, H.~Anahideh, J.~M. Rosenberger, D.~Martinez, V.~C. Chen, and B.~P.
  Wang, ``Global optimization of non-convex piecewise linear regression
  splines,'' \emph{Journal of Global Optimization}, vol.~68, no.~3, pp.
  563--586, 2017.

\bibitem{Elmachtoub:2021}
A.~N. Elmachtoub and P.~Grigas, ``Smart ``predict, then optimize'',''
  \emph{Management Science}, 2021.

\bibitem{Misic:2017}
V.~V. {Mi{\v{s}}i{\'c}}, ``{Optimization of Tree Ensembles},'' \emph{arXiv
  e-prints}, p. arXiv:1705.10883, May 2017.

\bibitem{Huchette:2020}
J.~{Huchette}, H.~{Lu}, H.~{Esfandiari}, and V.~{Mirrokni}, ``{Contextual
  Reserve Price Optimization in Auctions via Mixed-Integer Programming},''
  \emph{arXiv e-prints}, p. arXiv:2002.08841, Feb. 2020.

\bibitem{Say:2017}
B.~Say, G.~Wu, Y.~Q. Zhou, and S.~Sanner, ``Nonlinear hybrid planning with deep
  net learned transition models and mixed-integer linear programming,'' in
  \emph{Proceedings of the Twenty-Sixth International Joint Conference on
  Artificial Intelligence, {IJCAI-17}}, 2017, pp. 750--756.

\bibitem{Katz:2020}
J.~Katz, I.~Pappas, S.~Avraamidou, and E.~N. Pistikopoulos, ``The integration
  of explicit mpc and relu based neural networks,'' \emph{IFAC-PapersOnLine},
  vol.~53, no.~2, pp. 11\,350--11\,355, 2020.

\bibitem{Zhang:2020}
Y.~Zhang, C.~Chen, G.~Liu, T.~Hong, and F.~Qiu, ``Approximating trajectory
  constraints with machine learning--microgrid islanding with frequency
  constraints,'' \emph{IEEE Transactions on Power Systems}, vol.~36, no.~2, pp.
  1239--1249, 2020.

\bibitem{Zhang:2021}
Y.~{Zhang}, H.~{Cui}, J.~{Liu}, F.~{Qiu}, T.~{Hong}, R.~{Yao}, and F.~{Li},
  ``{Encoding Frequency Constraints in Preventive Unit Commitment Using Deep
  Learning with Region-of-Interest Active Sampling},'' \emph{arXiv e-prints},
  p. arXiv:2102.09583, Feb. 2021.

\bibitem{Amos:2016}
B.~{Amos}, L.~{Xu}, and J.~{Zico Kolter}, ``{Input Convex Neural Networks},''
  \emph{arXiv e-prints}, p. arXiv:1609.07152, Sep. 2016.

\bibitem{Chen:2020}
Y.~Chen, Y.~Shi, and B.~Zhang, ``Data-driven optimal voltage regulation using
  input convex neural networks,'' \emph{Electric Power Systems Research}, vol.
  189, p. 106741, 2020.

\bibitem{Misyris:2021}
G.~S. {Misyris}, J.~{Stiasny}, and S.~{Chatzivasileiadis}, ``{Capturing Power
  System Dynamics by Physics-Informed Neural Networks and Optimization},''
  \emph{arXiv e-prints}, p. arXiv:2103.17004, Mar. 2021.

\bibitem{Bastin:2019}
B.~Bastin, ``Dynamics and control of a future off-shore wind power hub,'' Ph.D.
  dissertation, M. Sc. thesis, University of Li{\`e}ge, 2019.

\bibitem{kingma_adam_2017}
D.~P. Kingma and J.~Ba, ``Adam: A method for stochastic optimization,''
  \emph{{arXiv}:1412.6980}, 2017, version: 9.

\bibitem{Tosatto:2021}
A.~Tosatto, M.~Dijokas, T.~Weckesser, S.~Chatzivasileiadis, and R.~Eriksson,
  ``\BIBforeignlanguage{English}{Sharing reserves through hvdc: Potential cost
  savings in the nordic countries},'' \emph{\BIBforeignlanguage{English}{IET
  Generation, Transmission and Distribution}}, vol.~15, no.~3, pp. 480--494,
  2021.

\bibitem{pytorch:2019}
A.~Paszke, S.~Gross \emph{et~al.}, ``Pytorch: An imperative style,
  high-performance deep learning library,'' in \emph{Advances in Neural
  Information Processing Systems 32}, H.~Wallach \emph{et~al.}, Eds.\hskip 1em
  plus 0.5em minus 0.4em\relax Curran Associates, Inc., 2019, pp. 8024--8035.

\bibitem{hart2011pyomo}
W.~E. Hart, J.-P. Watson, and D.~L. Woodruff, ``Pyomo: modeling and solving
  mathematical programs in python,'' \emph{Mathematical Programming
  Computation}, vol.~3, no.~3, pp. 219--260, 2011.

\bibitem{bynum2021pyomo}
M.~L. Bynum, G.~A. Hackebeil, W.~E. Hart, C.~D. Laird, B.~L. Nicholson, J.~D.
  Siirola, J.-P. Watson, and D.~L. Woodruff, \emph{Pyomo--optimization modeling
  in python}, 3rd~ed.\hskip 1em plus 0.5em minus 0.4em\relax Springer Science
  \& Business Media, 2021, vol.~67.

\bibitem{gurobi}
\BIBentryALTinterwordspacing
{Gurobi Optimization, LLC}, ``{Gurobi Optimizer Reference Manual},'' 2022.
  [Online]. Available: \url{https://www.gurobi.com}
\BIBentrySTDinterwordspacing

\bibitem{wandb}
\BIBentryALTinterwordspacing
L.~Biewald, ``Experiment tracking with weights and biases,'' 2020, software
  available from wandb.com. [Online]. Available: \url{https://www.wandb.com/}
\BIBentrySTDinterwordspacing

\bibitem{DTU_DCC_resource}
\BIBentryALTinterwordspacing
{DTU Computing Center}, ``{DTU Computing Center resources},'' 2021. [Online].
  Available: \url{https://doi.org/10.48714/DTU.HPC.0001}
\BIBentrySTDinterwordspacing

\bibitem{paleyes_challenges_2022}
\BIBentryALTinterwordspacing
A.~Paleyes, R.-G. Urma, and N.~D. Lawrence, ``Challenges in deploying machine
  learning: A survey of case studies,'' \emph{ACM Comput. Surv.}, apr 2022.
  [Online]. Available: \url{https://doi.org/10.1145/3533378}
\BIBentrySTDinterwordspacing

\end{thebibliography}

\end{document}